\documentstyle[aps,epsf]{revtex}

\newcommand{\troisj}[6]{\left( \begin{array}{ccc} #1&#2&#3\\#4&#5&#6\end{array}\right)}
\newcommand{\asixj}[6]{A \left\{\begin{array}{ccc} #1&#2&#3\\#4&#5&#6 \end{array}\right\}}
\newcommand{\sixj}[6]{\left\{\begin{array}{ccc} #1&#2&#3\\#4&#5&#6 \end{array}\right\}}
\newcommand{\aneufj}[9]{A \left\{\begin{array}{ccc} #1&#2&#3\\#4&#5&#6\\#7&#8&#9 \end{array}\right\}}
\newcommand{\neufj}[9]{\left\{\begin{array}{ccc} #1&#2&#3\\#4&#5&#6\\#7&#8&#9 \end{array}\right\}}

\begin{document}

\draft
\title{Solutions of the Faddeev-Yakubovsky equations for the four nucleons scattering states}
\author{F. Ciesielski and J. Carbonell}
\address{Institut des Sciences Nucl\'{e}aires,
        53, Av. des Martyrs, 38026 Grenoble, France}
\date{\today}
\maketitle

\begin{abstract}
The Faddeev-Yakubowsky equations in configuration space have been solved for
the four nucleon system. The results with an S-wave interaction model
in the isospin approximation are presented. They concern the bound
and scattering states below the first three-body threshold.
The elastic phase-shifts for the N+NNN reaction in different ($S,T$) channels are given and
the corresponding low energy expansions are discussed.
Particular attention is payed to the n+t elastic cross section. Its resonant structure
is well described in terms of a simple NN interaction.
First results concerning the S-matrix for the coupled N+NNN-NN+NN channels and
the strong deuteron-deuteron scattering length are obtained.
\end{abstract}

\section{Introduction}

\begin{figure}[hbtp]
\begin{center}\epsfxsize=12cm\epsfysize=12cm\mbox{\epsffile{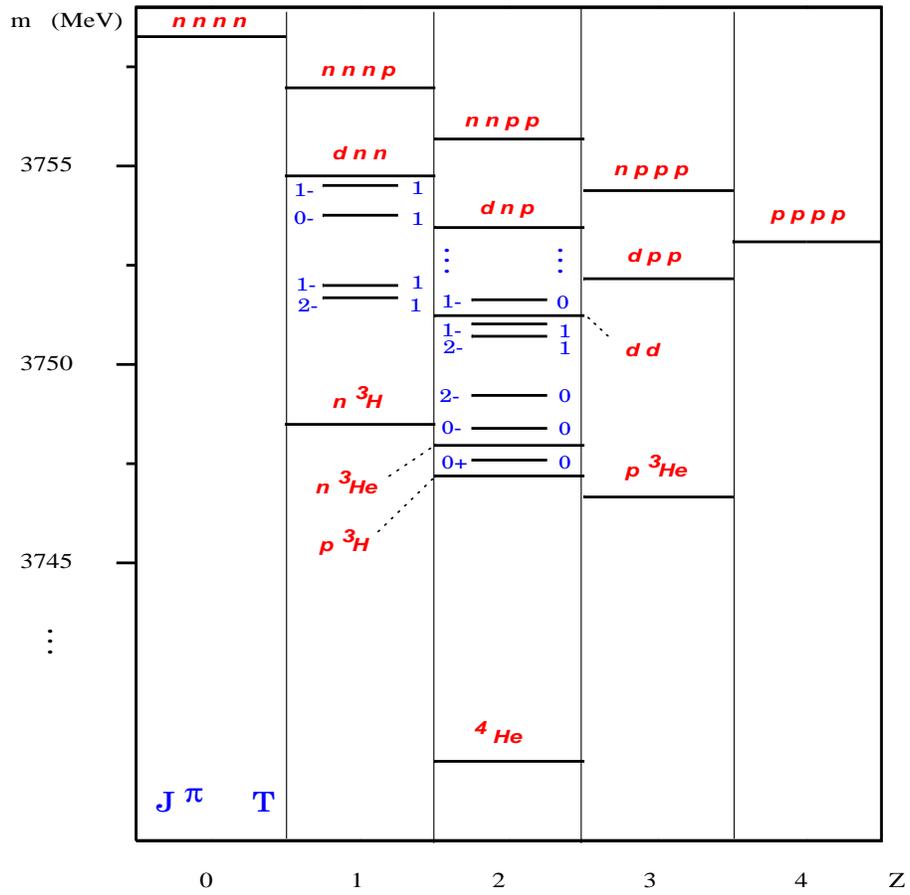}}\end{center}
\caption{The A=4 chart with the more relevant thresholds and resonances $(J^{\pi},T)$.
The vertical axis represents a mass scale; the horizontal one distinguishes
the different values of the electric charge $Z$.}\label{figchart}
\end{figure}

The four nucleon bound state calculations have, in the last years, reached a high level of accuracy and consistency
at least as far as the solutions of the corresponding equations are concerned \cite{SSK92,GK93,GWKHG95,PISA95}.
This situation contrasts with the 4N scattering problem 
where despite some pioneering and relevant results \cite{TJON76,F84,F89,F94,UOT93,UOT95,APS91,SCW95,YAKOVLEV95,HOF97},
there is a manifest lack of convergence among the different groups and methods even when using simple interactions.
This problem is not only a general extension of the 3-body one, in the sustained task
of the nuclear Few-Body community to deal with increasingly complex systems,
but we believe it constitutes a qualitative jump in our understanding of nuclear systems.
Indeed the continuum spectrum of the 4N system (see Figure~\ref{figchart}),
with its rich variety of thresholds and structures,
provides a bridge between the relative simplicity of the A=2,3 problems and the complexity of many-body systems.
Even when restricted to the energies below the first 3-body breakup threshold,
the presence of several resonances at each Z channel, the existence of the almost
degenerate p+t and n+$^3$He thresholds with,
in the middle, the first $0^+$ excitation of the $^4$He ground state
make the understanding of the A=4 chart in terms of fundamental NN interactions
an exciting and redoubtable theoretical challenge.

We present here the first solution of the Faddeev-Yakubovsky (FY) equations in configuration space 
for the four nucleon scattering problem.
Although  the results concerning the bound states ($^4$He and $^4$He$^*$)
will be discussed in some detail, our main interest lies 
in the 4N continuum spectrum, i.e. the N+NNN elastic scattering and its coupling
to the first inelastic NN+NN threshold.

The resolution method is based on the angular momentum expansion of the FY amplitudes
and the spline expansion of their radial parts.
Orthogonal collocation is used to generate a linear system which
is solved by iterative procedures. The scattering observables are extracted from a direct inspection of 
the FY amplitudes in the asymptotic region, in a natural extension of the methods developed
for the three-body case in \cite{CGM93}.

The results presented in this paper have all been obtained by using an S-wave
NN interaction model and the isospin symmetry hypothesis.
The Coulomb and mass difference effects are thus not included.
This choice, guided by methodological reasons, allows a presentation of the formalism and methods
in a relatively simple framework.
Is is remarkable, however, that such a simple model provides a very good description of low
energy scattering observables even if, as in the n+t case, they are not totally trivial.
Some first results including realistic interactions have already been
reported elsewhere \cite{CCG97} and will be the subject of subsequent publications.

The paper is organized as follows.
In the next section we describe the general formalism and
the simplifications arising in the case of 4 identical particles.
This section contains also the spin, isospin and angular momentum algebra.
In Section 3 we give some details of the numerical methods used.
In Section 4 the results will be presented.
They include the $^4$He ground and first excited state, the elastic phase-shifts
and low energy parameters
for the N+NNN reaction and the N+NNN~$\rightarrow$~NN+NN first inelastic channel.
The energies are restricted below the 3- and consequently 4-body break-up.
Conclusions and perspectives will be given in the last section.

\section{The Formalism}

\subsection{Faddeev-Yakubovsky equations}
\begin{figure}[hbtp]
\begin{center}\epsfxsize=12cm\mbox{\epsffile{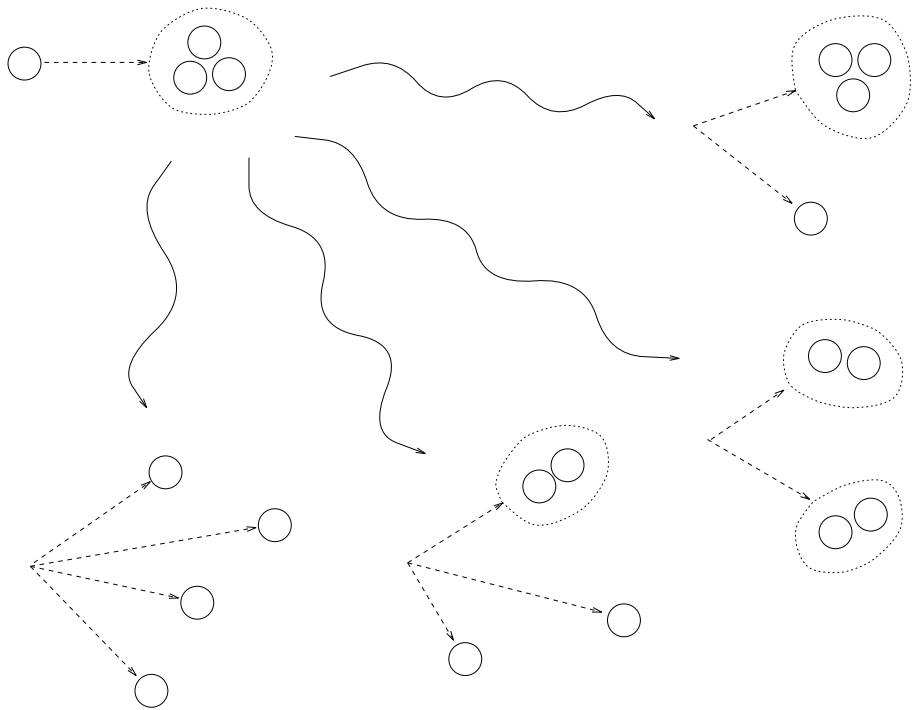}}\end{center}
\caption{Different asymptotics to be accounted for in a 1+3 collision.}\label{VOIES}
\end{figure}

With the aim of solving the Schr\"odinger equation for N particles
interacting via a pairwise potential $V_{ij}$
\begin{equation}\label{SCH}
(E-H_0) \Psi=  \sum_{i<j}V_{ij}  \Psi  
\end{equation}
Yakubovsky \cite{YAKU67}, generalizing Faddeev's work for N=3 \cite{FAD60,FAD61},
wrote a set of equations
whose solutions verify (\ref{SCH}) and which provides a proper mathematical scheme
to account for the variety of physical situations involved (see, e.g. Figure \ref{VOIES}).
In the N=4 case, the FY equations can be obtained by first splitting the total wavefunction $\Psi$
in the usual Faddeev amplitudes, $\Psi_{ij}$, associated with each interacting pair:
\[ \Psi= \sum_{i<j}\Psi_{ij} = \Psi_{12}+\Psi_{13}+\Psi_{14}+\Psi_{23}+\Psi_{24}+\Psi_{34},\]
and requiring them to be a solution of the system of coupled equations 
\begin{equation}\label{FED}
(E-H_0)\Psi_{ij}=V_{ij} \sum_{k<l}\Psi_{kl}
\end{equation}
or equivalently in its integral form
\begin{equation}\label{FEI}
 \Psi_{ij}=G_0V_{ij} \Psi  
\end{equation}
with
\[ G_0=(E-H_0)^{-1}\]

Each amplitude $\Psi_{ij}$ is in its turn split in 3 parts, the FY amplitudes,
corresponding to the different asymptotics of the remaining two particles:
\begin{equation}\label{deffy}
\Psi_{ij}=  \Psi_{ij,k}^l+\Psi_{ij,l}^k+\Psi_{ij,kl}\qquad i<j,k<l
\end{equation}
and obeying the following system of coupled equations : 
\begin{eqnarray}\label{FYDE}
(E-H_0-V_{ij})\Psi_{ij,k}^l&=&V_{ij}\left(
\Psi_{ik,j}^l+\Psi_{ik,l}^j+\Psi_{ik,lj} + \Psi_{jk,i}^l+\Psi_{jk,l}^i+\Psi_{jk,il} \right) \cr
(E-H_0-V_{ij})\Psi_{ij,l}^k&=&V_{ij}\left(
\Psi_{il,j}^k+\Psi_{il,k}^j+\Psi_{il,kj} + \Psi_{jl,i}^k+\Psi_{jl,k}^i+\Psi_{jl,ik}\right)\cr
(E-H_0-V_{ij})\Psi_{ij,kl}&=&V_{ij}\left(\Psi_{kl,i}^j+\Psi_{kl,j}^i+\Psi_{kl,ij}\right)  
\end{eqnarray}
in which an amplitude $\Psi_{\alpha>\beta,\gamma}^{\delta}$, not defined by (\ref{deffy}),
has to be understood as being identical to $\Psi_{\beta\alpha,\gamma}^{\delta}$.
Any solution of this system of 18 coupled equations, called the FY equations,
is a solution of (\ref{FED}) and consequently of the initial problem (\ref{SCH}). 
Its advantage lies in the possibility to define for system (\ref{FYDE}) 
appropriate boundary conditions ensuring the unicity of the solution.
Indeed when one of the particle, e.g. labeled by $l$, is out of reach of the interaction, 
all the amplitudes in (\ref{FYDE}) tend to zero except $\Psi_{ij,k}^l$ and circular
permutations on $ijk$ which obey
\begin{equation}\label{FYA1}
(E-H_0-V_{ij})\Psi_{ij,k}^l=V_{ij}\left(\Psi_{ik,j}^l+ \Psi_{jk,i}^l \right) 
\end{equation}
This system of equations, resulting from (\ref{FYDE}),
is equivalent to the 3N Faddeev equations for the particles $(ijk)$.
In a similar way, when the $(ij)$ and $(kl)$ clusters are free from interaction the only non
vanishing amplitudes are $\Psi_{ij,kl}$ and $\Psi_{kl,ij}$ and
their corresponding equations in (\ref{FYDE}) tends to
\begin{equation}\label{FYA2}
(E-H_0-V_{ij})\Psi_{ij,kl}=V_{ij}\Psi_{kl,ij}  
\end{equation}

It is worth noticing that the FY amplitudes can be written
in terms of the Faddeev amplitudes in the form:
\begin{eqnarray}\label{FY1}
\Psi_{ij,k}^l&=& G_{ij} V_{ij} \; \left( \Psi_{ik}+\Psi_{jk}  \right) \cr
\Psi_{ij,kl} &=& G_{ij} V_{ij} \; \Psi_{kl}   
\end{eqnarray}
where
\[ G_{ij}=(E-H_0-V_{ij})^{-1} \]
and, according to (\ref{FEI}), in terms of the total wavefunction:
\begin{eqnarray}\label{FY2}
\Psi_{ij,k}^l &=& G_{ij} \; V_{ij} \;G_0  \;\left( V_{ik} + V_{jk}  \right) \Psi \cr
\Psi_{ij,kl}  &=& G_{ij} \; V_{ij} \;G_0  \; V_{kl} \Psi    
\end{eqnarray}

Disregarding the internal degrees of freedom (like spin, isospin),
the natural basis for the configuration space is provided
by the positions of the different particles
\begin{equation}\label{baser}
\mid\vec{r}_1\vec{r}_2\vec{r}_3\vec{r}_4>=
\mid\vec{r}_1>\otimes\mid\vec{r}_2>\otimes\mid\vec{r}_3>\otimes\mid\vec{r}_4>
\end{equation}

In order to remove the center of mass motion,
it is useful to introduce the relative Jacobi coordinates.
Two sets of Jacobi coordinates can be defined for each of the 4! arrangements $(ijkl)$.
One of K type:
\begin{eqnarray}\label{K}
\vec{x}_K(ijkl)&=&\sqrt{2\mu_{i,j}\over m}(\vec{r}_j-\vec{r}_i)\cr
\vec{y}_K(ijkl)&=&\sqrt{2\mu_{ij,k}\over m}\left(\vec{r}_k-{m_i\vec{r}_i+m_j\vec{r}_j\over m_{i}+m_{j}}\right)\cr 
\vec{z}_K(ijkl)&=&\sqrt{2\mu_{ijk,l}\over m}
\left(\vec{r}_l-{m_i\vec{r}_i+m_j\vec{r}_j+m_k\vec{r}_k\over m_{i}+m_{j}+m_{k}}\right)
\end{eqnarray}
and one of H type:
\begin{eqnarray}\label{H}
\vec{x}_H(ijkl)&=&\sqrt{2\mu_{i,j}\over m}(\vec{r}_j-\vec{r}_i)\cr
\vec{y}_H(ijkl)&=&\sqrt{2\mu_{k,l}\over m}(\vec{r}_l-\vec{r}_k)\cr
\vec{z}_H(ijkl)&=&\sqrt{2\mu_{ij,kl}\over m} 
                     \left({m_k\vec{r}_k+m_l\vec{r}_l\over m_{k}+m_{l}} -
                     {m_i\vec{r}_i+m_j\vec{r}_j\over m_{i}+m_{j}}\right)
\end{eqnarray}
in which $m$ is an arbitrary mass taken as a reference, and $\mu_{\alpha,\beta}$ is the reduced mass
of clusters $\alpha$ and $\beta$.
However some of these 48 coordinate sets are redundant. For instance those obtained
by exchanging $i \leftrightarrow j$ in a K-set or $i \leftrightarrow j$
or/and $k \leftrightarrow l$ in the H-set are equivalent.
This yields 18 (12K+6H) arbitrary and physically non equivalent Jacobi sets,
as many as FY amplitudes.
Any of these coordinate sets, suitably completed with the center of mass coordinate $\vec{R}$,
constitutes an equivalent description of the 4 particles configuration space.
That provides 18 coordinate sets and the corresponding bases for the configuration space,
equivalent to (\ref{baser}), that will be written in the form 
\(\mid\vec{x}_K\vec{y}_K\vec{z}_K(ijkl)\vec{R}>\) or
\(\mid\vec{x}_H\vec{y}_H\vec{z}_K(ijkl)\vec{R}>\).
The degrees of freedom related to the center of mass motion separate in nonrelativistic
dynamics and will be hereafter omitted.
Although each FY amplitude could be in principle expressed in terms
of any of these bases, only one of them is appropriate for expanding it.
We will denote the resulting components by:
\begin{eqnarray*}
\Phi_{ij,k}^l(\vec{x},\vec{y},\vec{z}) &\equiv&
<\vec{x}_K\vec{y}_K\vec{z}_K(ijkl)|\Phi_{ij,k}^l>\cr
\Phi_{ij,kl} (\vec{x},\vec{y},\vec{x}) &\equiv&
<\vec{x}_H\vec{y}_H\vec{z}_H(ijkl)|\Phi_{ij,kl} >  
\end{eqnarray*}
The bases described above have to be completed to account
for other degrees of freedom like spin, isospin, etc.
Further details about the formalism and the relation between the different bases sets can be found,
e.g., in \cite{Fred97}.

\subsection{Identical particles}

In the case of four identical particles, the 18 FY amplitudes can be obtained by the action of the transposition 
permutation operators $P_{ij}$ on two of them,
arbitrarily chosen provided that one is of K type and the other one of H type.
We have taken  \(K\equiv\Psi_{12,3}^4\) and \(H\equiv\Psi_{12,34}\). 
The 4-body problem is solved by determining the two $K,H$ amplitudes
which satisfy the following equations:
\begin{eqnarray}
(E-H_0-V) K&=&V\left[ (P_{23}+P_{13})\;(\varepsilon + P_{34})\; K
+\varepsilon(P_{23}+P_{13})\; H\right]\label{FYE1}\\
(E-H_0-V) H&=&V\left[ (P_{13}P_{24}+P_{14}P_{23}) \; K +
P_{13}P_{24} \; H\right]  \label{FYE2}
\end{eqnarray}
where $\varepsilon=\pm1$ depending on whether the particles are bosons or fermions.
The asymptotic equations, i.e. the equivalent of (\ref{FYA1},\ref{FYA2}) are in this case
\begin{eqnarray}\label{FYAI1}
(E-H_0-V) K&=&\varepsilon \;V (P_{23}+P_{13}) K \\
(E-H_0-V) H&=&V	P_{13}P_{24} \; H  \label{FYAI2}
\end{eqnarray}
The total wave function is then given by:
\begin{eqnarray}
     \Psi  &=&  \Psi_{1+3} + \Psi_{2+2} \cr  
\Psi_{1+3} &=&  \left[ 1+ \varepsilon(P_{13}+P_{23}       ) \right]\;
                \left[ 1+ \varepsilon(P_{14}+P_{24}+P_{34}) \right]K  \label{psi13}\\
\Psi_{2+2} &=&  \left[ 1+\varepsilon( P_{13}+ P_{23}+ P_{14}+ P_{24} ) + P_{13}P_{24}\right] \;H  \label{psi22}
\end{eqnarray}
Each amplitude $\Phi=K,H$ is considered as a function of its natural set of Jacobi coordinates $\vec{x}_{\Phi},\vec{y}_{\Phi},\vec{z}_{\Phi}$, 
defined respectively by equations (\ref{K}) and (\ref{H}) with $(ijkl)=(1234)$ and $m=m_i$:
\begin{eqnarray*}
\vec{x}_K&=& \vec{r}_2-\vec{r}_1 \cr
\vec{y}_K&=&\displaystyle \sqrt{4 \over3}\left(\vec{r}_3-{\vec{r}_1+\vec{r}_2\over2}\right)\cr
\vec{z}_K&=&\displaystyle \sqrt{3 \over2}\left(\vec{r}_4-{\vec{r}_1+\vec{r}_2+\vec{r}_3\over3}\right)
\end{eqnarray*}\qquad
\begin{eqnarray*}
\vec{x}_H&=& \vec{r}_2-\vec{r}_1 \cr
\vec{y}_H&=& \vec{r}_4-\vec{r}_3 \cr
\vec{z}_H&=&\displaystyle \sqrt{2}\left({\vec{r}_3+\vec{r}_4\over2}-{\vec{r}_1+\vec{r}_2\over2}\right)
\end{eqnarray*}

\begin{figure}[hbtp]
\begin{center}
\epsfxsize=8cm
\mbox{\epsffile{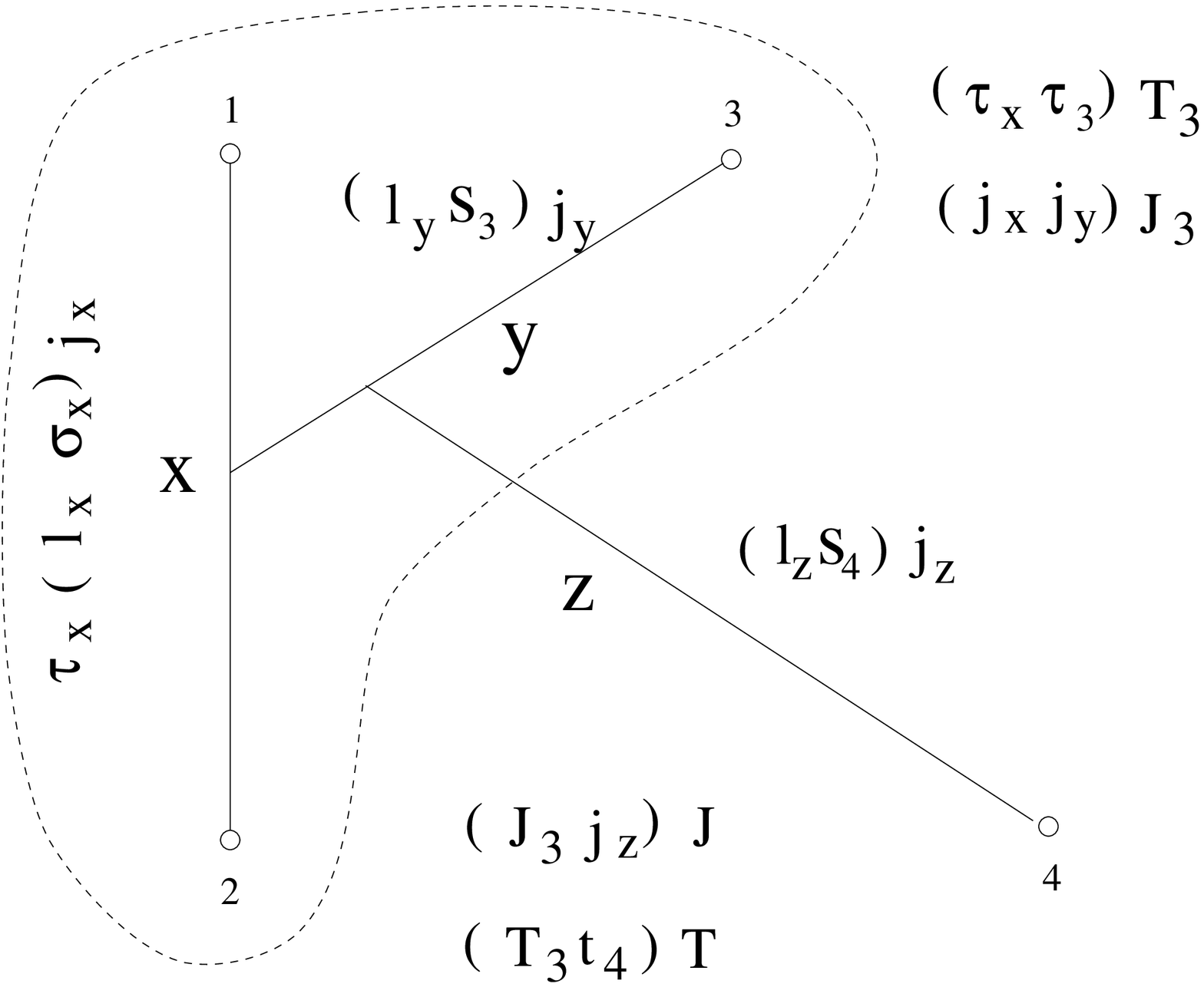}}
\epsfxsize=8cm
\mbox{\epsffile{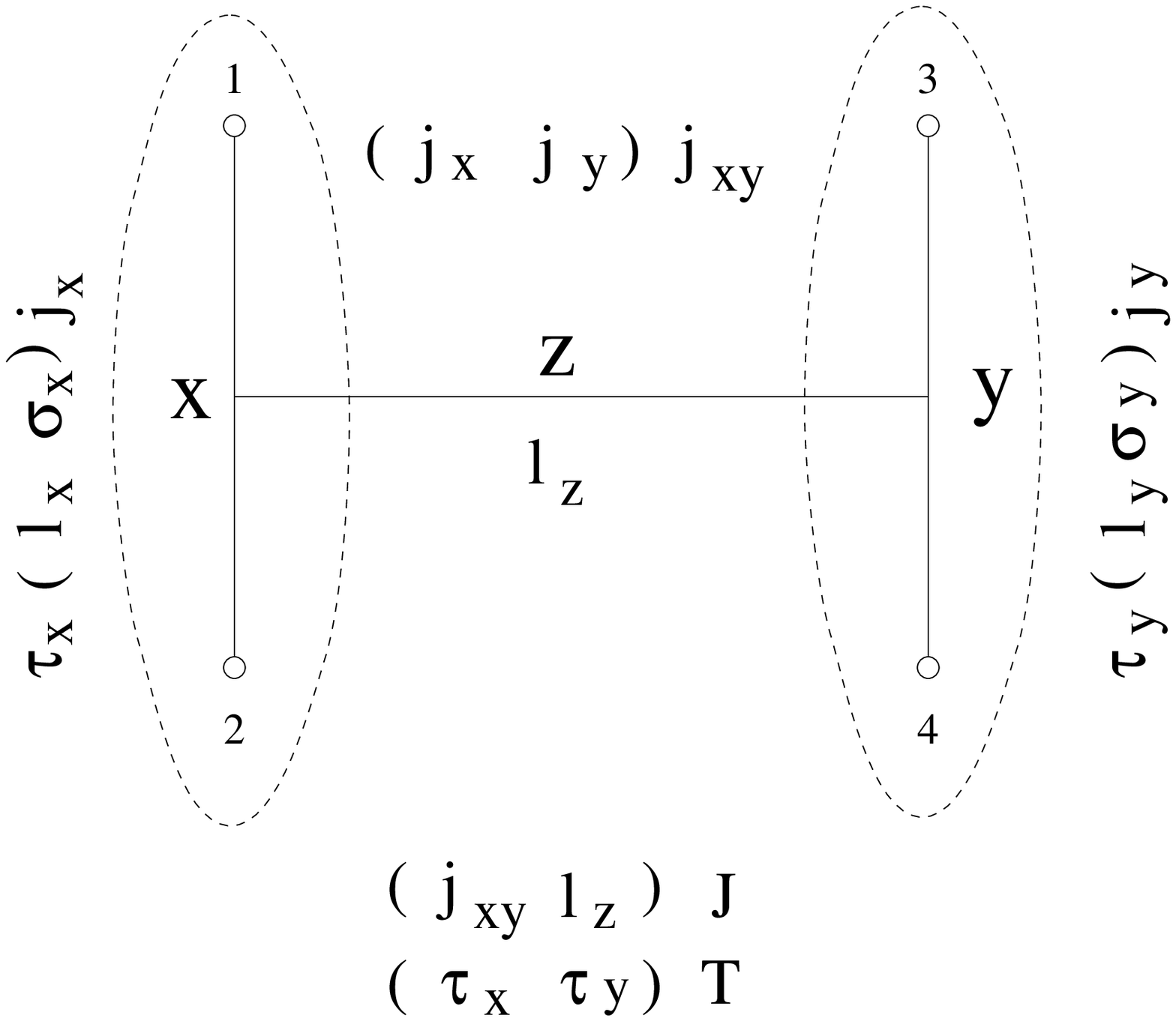}}
\end{center}
\caption{Spin, isospin and angular momentum coupling schemes used for the K and H
Faddeev-Yakubovsky amplitudes.}\label{coupling}
\end{figure}

They are expanded in angular momentum variables for each coordinate according to
\begin{equation}\label{KPW}
<\vec{x}\vec{y}\vec{z}|{\Phi}>=
\sum_{\alpha} \; {\phi_{\alpha}(x,y,z)\over xyz} \; Y_{\alpha} (\hat{x},\hat{y},\hat{z}) 
\end{equation} 
where $Y_{\alpha}$ are generalized tripolar harmonics containing spin, isospin and angular momentum variables
and the functions $\phi_{\alpha}$, the reduced radial FY components, are the unknowns.
The label $\alpha$ represents the set of intermediate quantum numbers defined in a given coupling scheme
and includes the specification for the type of amplitudes (K or H). We have used the following couplings,
represented in Figure~\ref{coupling}:
\begin{eqnarray}\label{devcoupling}
\mbox{K amplitudes}  &:&\left\{ \left[ (t_1 t_2)_{\tau_x} t_3 \right]_{T_3} t_4 \right\}_T \otimes
\left\{ \left[ \left( l_x (s_1 s_2)_{\sigma_x} \right)_{j_x} (l_y s_3)_{j_y}
\right]_{J_3} (l_z s_4)_{j_z} \right\}_{J^{\pi}}     \cr
\mbox{H amplitudes}  &:& \left[ (t_1 t_2)_{\tau_x} (t_3 t_4)_{\tau_y} \right]_T \otimes
\left\{ \left[ \left( l_x (s_1 s_2)_{\sigma_x} \right)_{j_x} 
\left( l_y (s_3 s_4)_{\sigma_y} \right)_{j_y} \right]_{j_{xy}} l_z \right\}_{J^{\pi}} 
\end{eqnarray}
where  $s_i$ and $t_i$ are the spin and isospin of the individual particles 
and $(J^{\pi},T)$ are respectively the total angular momentum, parity, and isospin of the 4-body system.
Each component $\phi_{\alpha}$ is thus labeled  by a set of 12 quantum numbers
to which the symmetry properties of the wavefunction impose the additional constraints:
$(-1)^{\sigma_x+\tau_x+l_x}=\varepsilon$ for K and $(-1)^{\sigma_x+\tau_x+l_x}=(-1)^{\sigma_y+\tau_y+l_y}=\varepsilon$ for H.
The total parity $\pi$ is given by $(-)^{l_x+l_y+l_z}$ in both coupling schemes.

The radial equations for the components $\phi_{\alpha}$ are obtained by projecting
each of equations (\ref{FYE1}-\ref{FYE2}) in its natural configuration space basis $\mid\vec{x}_{\Phi},\vec{y}_{\Phi},\vec{z}_{\Phi}>$.
Several steps further \cite{Fred97} we end with a system of coupled integro-differential equations 
which, most generally, can be written in the form:
\begin{eqnarray}
\sum_{\alpha'} \hat{D}_{\alpha\alpha'} \phi_{\alpha'}(x,y,z)=
\sum_{\alpha'}V_{\alpha\alpha'}(x)\sum_{\alpha"}\bigg[ &&
   f_{\alpha'\alpha"} \; 
   \phi_{\alpha"}(x^f_{\alpha'\alpha"},y^f_{\alpha'\alpha"},z^f_{\alpha'\alpha"}) \cr
&+&\int_{-1}^{+1} du \; h_{\alpha'\alpha"}(x,y,z,u) \; 
   \phi_{\alpha"}(x^h_{\alpha'\alpha"},y^h_{\alpha'\alpha"},z^h_{\alpha'\alpha"}) \cr
&+&\int_{-1}^{+1} du \;\int_{-1}^{+1} dv \; g_{\alpha'\alpha"}(x,y,z,u,v) \; 
   \phi_{\alpha"}(x^g_{\alpha'\alpha"},y^g_{\alpha'\alpha"},z^g_{\alpha'\alpha"})\bigg] \label{fse}
\end{eqnarray}
with
\begin{eqnarray*}
\hat{D}_{\alpha\alpha'}&=& (E+\frac{\hbar^2}{m}\Delta_{\alpha})\delta_{\alpha\alpha'}-V_{\alpha\alpha'}(x)\cr
\Delta_{\alpha}&=&\partial_x^2-\frac{l_x(l_x+1)}{x^2}+\partial_y^2-\frac{l_y(l_y+1)}{y^2}+\partial_z^2-\frac{l_z(l_z+1)}{z^2}
\end{eqnarray*}
The functions  $f_{\alpha'\alpha"},\; h_{\alpha'\alpha"},\; g_{\alpha'\alpha"}$ 
contain all the spin, isospin and angular momentum couplings.
The arguments $x^f_{\alpha'\alpha"},x^h_{\alpha'\alpha"},x^g_{\alpha'\alpha"},\ldots$ 
are functions of $(x,y,z,u,v)$ in the more general case, and are detailed in the Appendix.
The system of equations (\ref{fse}) has been explicitly written in \cite{MERKU84}
for the case of four identical bosons.  

The FY components for the different (S,T) channels in the S-wave approximation, i.e. with all orbital 
angular momenta in expansion (\ref{KPW}) equal to 0,
are listed in Table~\ref{tab_amplis}. In this table, the symbols ``~$\rightarrow$~'' and 
``~$\sim~$'' denote respectively the amplitudes corresponding
to an asymptotic N+NNN or NN+NN channel.

Note that, contrary to the 3N problem, the number of FY components appearing in (\ref{fse}) is infinite 
even when the pair interaction is restricted to a finite number of partial waves.
This divergence comes only from the existence of the $l_z$ additional degree of freedom in the K-like amplitudes.

\subsection{Boundary conditions}

For all the physical problems we have considered, the boundary conditions can be written in the Dirichlet form.
The use of reduced radial FY components $\phi_{\alpha}$ in (\ref{KPW})
imposes for any kind of solution the regularity conditions: 
\begin{equation}
\phi_{\alpha}(x,y,0)=\phi_{\alpha}(x,0,z)=\phi_{\alpha}(0,y,z)=0  
\end{equation}
For the bound state problem these conditions are completed by forcing the components $\phi_{\alpha}$
to vanish on the hypercube $[0,x_N]\times[0,y_N]\times[0,z_N]$, i.e:
\begin{equation}\label{bsbc}
\phi_{\alpha}(x,y,0)=\phi_{\alpha}(x,y,z_{N})=
\phi_{\alpha}(x,0,z)=\phi_{\alpha}(x,y_{N},z)=
\phi_{\alpha}(0,y,z)=\phi_{\alpha}(x_{N},y,z)=0  
\end{equation}
For the scattering problems
the boundary conditions are implemented by imposing at large enough values of $z$
the asymptotic behavior of the solutions.
Thus, for the N+NNN elastic case we impose at $z_N$ the solution of the 3N problem for all the quantum numbers
$\alpha_a$ corresponding to the open asymptotic channel $a$
\begin{equation}\label{asym1}
\phi_{\alpha_a}(x,y,z_N)= t_{\alpha_a}(x,y)
\end{equation}
where the functions $t_{\alpha_a}(x,y)$ are the Faddeev amplitudes of the 3N problem.
Indeed, at large values of $z$ and for energies below the first inelastic threshold,
the solution of equation (\ref{FYAI1}) 
factorizes into a bound state solution of the 3N Faddeev equations and
a plane wave in the $z$ direction with momentum $k_a$,
whereas the solution of (\ref{FYAI2}) vanishes. One then has, e.g. for an S-wave,
\begin{equation}\label{DLOGPHI}
\phi_{\alpha_a}(x,y,z)\sim t_{\alpha_a}(x,y) \sin(k_az+\delta_a)
\end{equation}
and the scattering observables are directly extracted from the logarithmic derivative of the 
K amplitude in the asymptotic region:
\begin{equation}\label{QCOTD}
k_a\; {\rm cot}(k_az+\delta_a)= {1\over\phi_{\alpha_a}(x,y,z) }\partial_z\phi_{\alpha_a} (x,y,z)
\end{equation}
Provided we are in the asymptotic domain, the phase shifts thus extracted have to be independent of $(x,y,z)$ 
and of the amplitude index $\alpha_a$. This provides a strong numerical test.
An additional advantage of this procedure is that it avoids any cumbersome multidimensional integrals.

In the presence of several open channels, like N+NNN and NN+NN e.g., several resolutions are needed.
The boundary conditions (\ref{asym1}) are simply generalized in the form
\begin{equation}\label{asym2}
\phi_{\alpha_a}(x,y,z_N)= \lambda_a \; f_{\alpha_a}(x,y)
\end{equation}
in which $\lambda_a$ are arbitrary real numbers
and the functions $f_{\alpha_a}$ coincide with $t_{\alpha_a}$ if $a$ is a N+NNN channel or
are analogous to the Faddeev amplitudes for (\ref{FYAI2}) if $a$ is a NN+NN channel.

Equation (\ref{asym1}) is generalized in the following way:
\begin{equation}\label{smatrix}
\varphi_{\alpha_{a}}(x,y,z)= \left(-\tau_{a}\;\hat{h}^{-}(k_{a}z) +
\sum_{a'}\tau_{a'}\;\sqrt{\frac{n_{a'}k_{a'}}{n_{a}k_{a}}}\; S_{aa'}\;
\hat{h}^{+}(k_{a}z) \right) \; f_{\alpha_a}(x,y)
\end{equation}
where $\hat{h}^{\pm}$ are the regularized Hankel functions \cite{TAYLOR},
and $n_{a}$ is a multiplicity number for the channel $a$ ($n_a=4$ for 1+3 and $n_a=3$ for 2+2 channels)
due to the symmetry properties of the total wave function (see, e.g. \cite{BENCZE95}).
The coefficients
$\tau$ (amplitudes of the incoming waves) and the S-matrix elements are the unknowns.
They are obtained by identifying with
the asymptotic form (\ref{smatrix}) the values $\phi_{\alpha_a},\partial_z\phi_{\alpha_a}$  at $z_N$ for
different solutions corresponding to different choices of $\lambda$'s
whose number equals the number of channels.
We remark that the momenta $k_a$ appearing in equations (\ref{DLOGPHI},\ref{QCOTD},\ref{smatrix})
are the conjugate variables of the $z$-Jacobi coordinates.
They are related to the center of mass kinetic energy $E_a$
of channel $a$ according to $E_a=\frac{\hbar^2 k_a^2}{m}$.
The physical momenta, conjugate to the physical intercluster distances, are $q_a=\sqrt{3 \over 2}k_a$ or $q_a=\sqrt{2}k_a$ depending 
whether $a$ corresponds to a 1+3 or 2+2 channel.

By the definition (\ref{smatrix}) one has
unitarity ($SS^{\dagger}=1$) and symmetry ($S_{ij}=S_{ji}$) relations.
Working with real solutions these properties are related (unitarity implies symmetry). However
none of them is a trivial consequence of the method used but a strong test of numerical accuracy. 

\begin{figure}[hbtp]
\begin{center}
{\small (a) elastic scattering at zero-energy~: $\phi_{\alpha}(x,y,z)$ asymptotically
 linear towards $z$}\\
\epsfxsize=7cm
\mbox{\epsffile{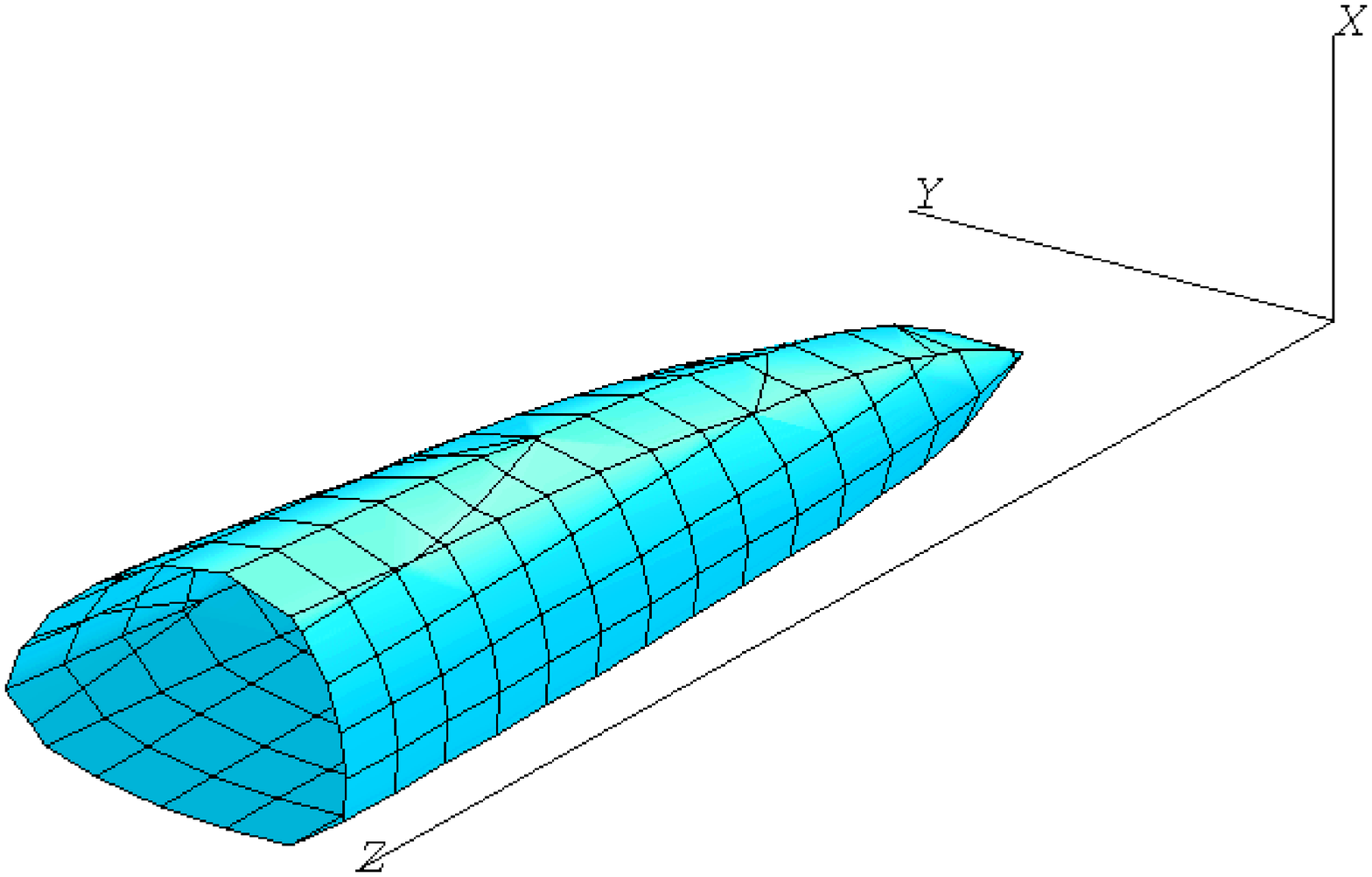}}
\epsfxsize=7cm
\mbox{\epsffile{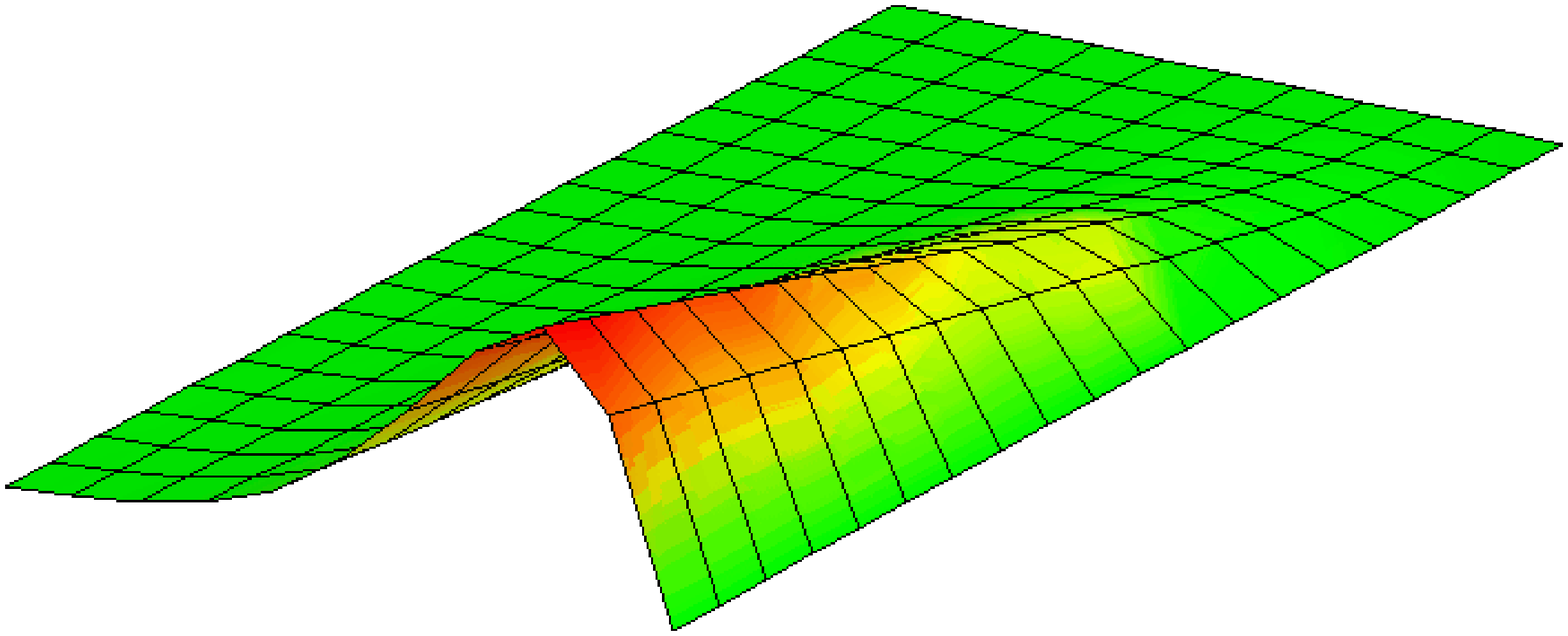}}
\\{\small (b) elastic scattering with positive kinetic energy~: $\phi_{\alpha}(x,y,z)$ asymptotically
 oscillating towards $z$}\\
\epsfxsize=7cm
\mbox{\epsffile{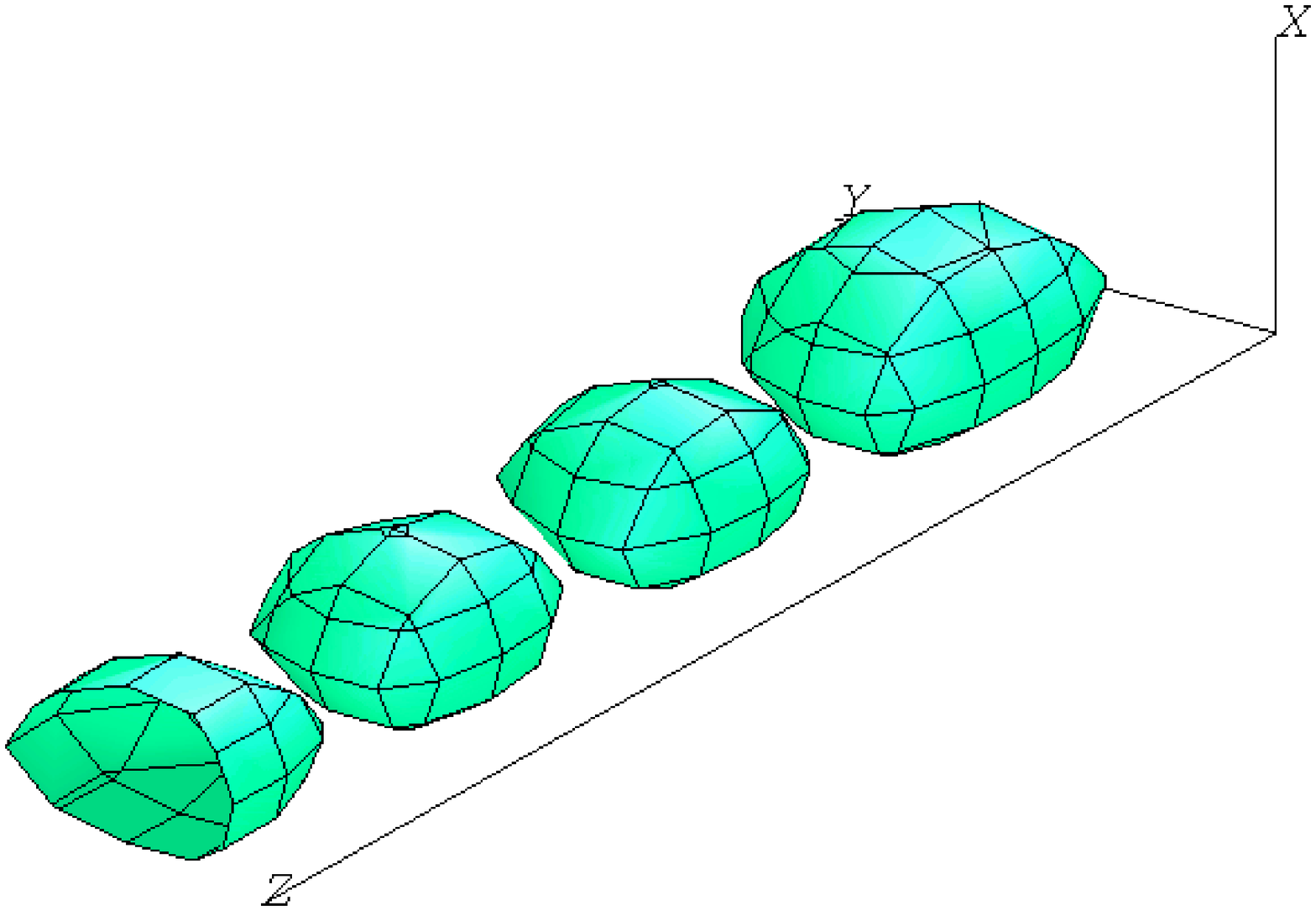}}
\epsfxsize=7cm
\mbox{\epsffile{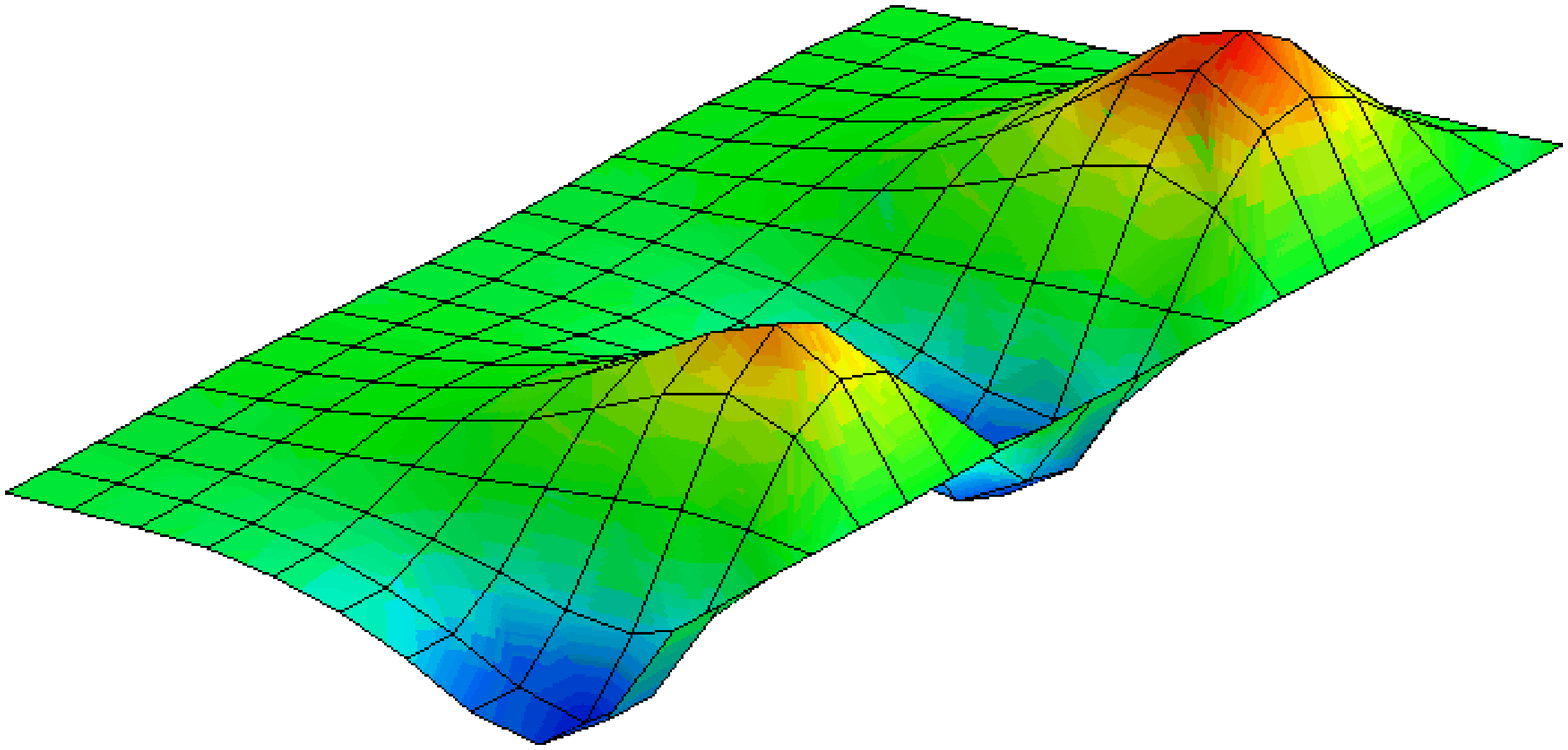}}
\end{center}
\caption{Representations of some asymptotically non-vanishing amplitudes for n+t elastic
scattering: (a)~zero-energy (b)~positive kinetic energy. On left, isosurfaces~; on right,
sections in $x$-variable. The asymptotic factorization between
an independent pattern on the $x,y$ 
coordinates and the $z$-variable motion appears clearly.}\label{REPRESNT}
\end{figure}

In order that the factorizability takes place the asymptotic 3N or 2N+2N states
have to be calculated with the same numerical scheme as that
used to solve the 4-body problem. This means in practice that
they are exact solutions of (\ref{FYAI1}) or (\ref{FYAI2}), once the z-dynamics are removed.
By doing so, the factorization property, valid only in cartesian coordinates, 
is an exact numerical property and leads to stable local results.
This behavior is illustrated in Figure~\ref{REPRESNT}
in which the FY amplitudes for an N+NNN elastic scattering
at $q = 0$ and $q > 0$ are represented.

\section{Numerical methods}

The numerical methods used are based on the Hermite spline expansion, orthogonal collocation \cite{PAYNE87}
and iterative procedures for solving the linear system.
An important step in their solution is the tensor trick \cite{SSK92,SKB89,SK90,S95}.

We look for the solutions $\phi_{\alpha}$ of the integro-differential system (\ref{fse}) in the form of a
tensor product of 1-dimensional cubic Hermite splines $S_i$:
\begin{equation}\label{sol1}
\phi_{\alpha}(x,y,z)=\sum_{i=0}^{2N_x+1}\sum_{j=0}^{2N_y+1}\sum_{k=0}^{2N_z+1} c_{\alpha ijk} S_i(x) S_j(y) S_k(z)
\end{equation}
defined on grids of non-equidistant $N_q+1$ points G$_q=\{q_0,q_1,\ldots,q_{N}\}$ where $q \equiv x,y,z$.
A grid G$_q$ will be defined by giving the number of intervals $N_q$, the end point $q_N$
and the constant scaling factor between two consecutive intervals, $A_q$.
We will use the following notation G$\equiv\{N_q,A_q,q_N\}$,
often extended to multi-domain grids according to
G$\equiv\{N_{q1},A_{q1},q_{N1};N_{q2},A_{q2},q_{N2}; \cdots\}$
 
The boundary conditions are easily implemented using the properties
of the spline functions (value and derivative equal to 0
or 1 at the grid points). They result in fixing 
the values of some of the unknown coefficients $c_{\alpha ijk}$ in the expansion (\ref{sol1}). 
By exchanging the indices of the two last spline functions in each variable the solution takes the form:
\begin{equation}\label{PHISPLINE}
\phi_{\alpha}(x,y,z)=
\sum_{i=1}^{2N_x} \; \sum_{j=1}^{2N_y} \; \sum_{k=1}^{2N_z} c_{\alpha ijk} \; S_i(x) S_j(y) S_k(z)
+ \sum_{i=1}^{2N_x}\sum_{j=1}^{2N_y} f_{\alpha ij} S_i(x) S_j(y) S_{2N_z+1}(z)
\end{equation}
where $f_{\alpha ij}$ are the coefficients of the asymptotic functions $f_{\alpha}(x,y)$ defined in  (\ref{asym2}).
In the cases of a closed channel or a bound state these coefficients are zero.

Collocation points associated with the 3-dimensional grid are used to generate a linear system,
the $c_{\alpha i j k}$ being the unknowns. The integral terms in  (\ref{fse}) are calculated with 
a gaussian quadrature rule with typically $N_u,N_v=6-12$ points in each angular variable $u$ and $v$.
In order to limit the number of parameters we have chosen $N_u=N_v=N_g$.

The number of unknowns is given by $N=8N_xN_yN_zN_c$ where $N_c=N_K+N_H$ is the number of FY components.
A rough estimation for the extreme cases of a four bosons bound state and of a scattering state
in a realistic problem leads to values 
$N_c=2-100$, $N_x,N_y=20-30$, $N_z=20-40$ and consequently to a number of
unknowns going from $N\sim 10^4$ to $N\sim 10^6$. This implies the use of a dense matrix
with $\sim 10^{12}$ coefficients.
Direct methods are not appropriate for solving such huge linear systems, and we have
used iterative techniques, which avoid any storage or inversion of the matrix.
The basic feature of any iterative method is to obtain the solution of the linear system
only by iterative application of the matrix to an initial vector.
This implies a complete calculation of the matrix elements each time it is necessary,
and requires the intensive use of parallel computers.  

In the case of scattering states, the boundary conditions (\ref{asym2}), responsible for
the second term in (\ref{PHISPLINE}),
generate a source term in (\ref{fse}) leading to a regular linear
system of the type $D(E)\vec{c}=\left[F+G+H\right]\vec{c}+\vec{b}$, where the different  
$D, F, G, H$ matrices are reminiscent of the $D_{\alpha \alpha'}, f_{\alpha \alpha'}, g_{\alpha \alpha'},
h_{\alpha \alpha'}$ operators. For brevity we will write the system in the form $A\vec{c}=\vec{b}$.

The numerical method we have chosen to solve this system is GMRES
(for Generalized Minimum RESidual algorithm) \cite{SAAD86}.
GMRES is a prototype of the so called Krilov subspaces projection methods.
Its aim is to minimize the residue $\vec{r}=\vec{b}-A\vec{c}$ of an approximate solution $\vec{c}$,
starting from a trial vector $\vec{c}_0$ and looking for its correction $\vec{c}-\vec{c}_0$ in the Krilov 
subspace ${\cal{K}}=\{\vec{r}_0, A\vec{r}_0, A^2\vec{r}_0,...,A^{p-1}\vec{r}_0\}$
such that the residue $\vec{r}$ is orthogonal to ${\cal{L}}=A{\cal{K}}$. 
When the dimension $p$ increases, the residue of the approximate solution $\vec{c}$
can be brought to an arbitrary small value, 
called tolerance. In most of the practical cases, a tolerance between $10^{-3}$ and $10^{-6}$ is sufficient.

\begin{figure}[hbtp]
\begin{center}
\epsfxsize=10cm\mbox{\epsffile{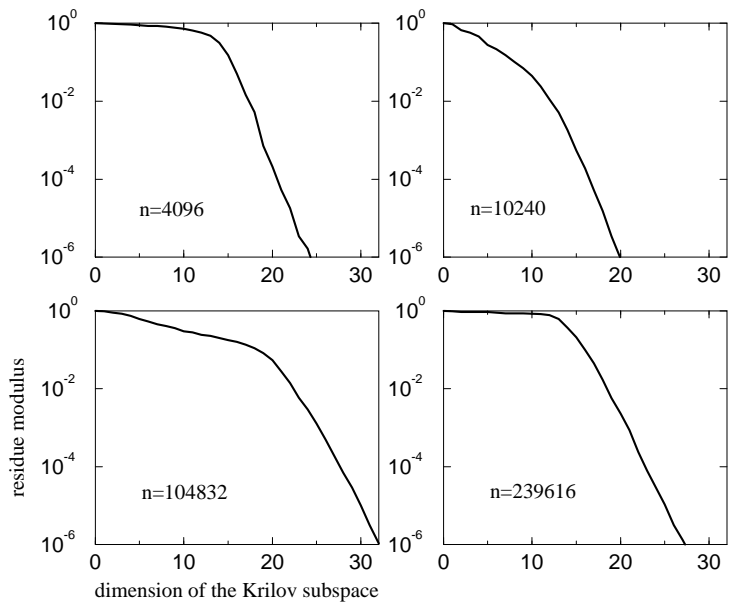}}
\end{center}
\caption{Residue modulus obtained with GMRES vs the dimension of the Krilov subspace,
i.e. the number of matrix applications,
for different numbers, n, of unknowns. The linear system $A\vec{c}=\vec{b}$ is normalized
such that $||\vec{b}||=1$. The initial guess is chosen to be $\vec{0}$.}\label{EVOLRES}
\end{figure}

GMRES is a powerful tool when the problem is well conditioned, what is almost never true in a realistic case. 
The way out is  to ``precondition'' the system, i.e. to solve the equivalent problem
$M^{-1}A\vec{c}=M^{-1}\vec{b}$ instead of $A\vec{c}=\vec{b}$.
The closer to $A$ is the matrix $M$, the better is the preconditioning.
Our choice was to take the matrix $M$ equal to $D$.
As pointed out in \cite{PAYNE87,SCHELL89,SSK92} its tensor structure, optimized by the choice
of cartesian coordinates, allows an easy inversion.
Our preconditioning technique gives us a converged result after $\simeq 30$ matrix applications,
for all the considered physical cases
and independently of the dimension of the matrix. Examples of convergence curves, i.e.
the evolution of the residue modulus at each step, are shown in Figure~\ref{EVOLRES}.

In the case of bound state, the asymptotic behavior of the wave function
and FY amplitudes leads to a singular system $D(E)\vec{c}=\left[F+G+H\right]\vec{c}$
that will be rewritten for convenience
in the form $A'\vec{c}=E\vec{c}$, the energy E being an eigenvalue.
It is well known that iterative methods are appropriate for extracting a few eigenvalues,
but only those with largest modulus.
With the 3-dimensional box-like boundary condition (\ref{bsbc}),
the existence of overwhelming highly positive eigenvalues
leaves no hope of obtaining directly the binding energy. 

A crafty trick is to solve the eigenproblem
$D^{-1}(E)\cdot\left[F+G+H\right]\vec{c}=\lambda\vec{c}$ where $E$ is now a parameter \cite{MT69,PAYNE87}. 
The value $E$ will be an eigenvalue of the initial problem when the spectrum $\{\lambda\}$ contains 1.
Furthermore, it can be shown by variational considerations
that the more excited the state (including the non physical
box states), the smaller the corresponding $\lambda$.
Thus the eigenvalues of physical interest can be obtained
with methods like IRA (Implicit Restarted Arnoldi algorithm) \cite{SAAD}.
We notice that the full inversion of $D$, including the two-body potential,
gives a better conditioned spectrum $\{\lambda\}$ than an inversion of its kinetic term alone,
and avoids some of the awkward negative eigenvalues generated by the repulsive part of the potential.
IRA is a generalization of the power method and gives the first
eigenvalue in $10-15$ matrix applications.
It is also based on Krilov subspaces, and approximates
the eigenvalues of $A'$ by those of the restriction of $A'$ to the
space spanned by $k$ vectors $\vec{x}_0, \vec{x}_1,...,\vec{x}_{k-1}$,
$\vec{x}_0$ being a trial arbitrary vector.
Nevertheless, this method requires several calculations for different values of the energy
and becomes numerically unstable when 
using highly repulsive two-body potentials, like the interatomic $^4$He-$^4$He one \cite{AZIZ3}.

A more robust technique was finally adopted, often referred as shift-invert method. 
The initial problem is written in the form $(A'-E_0)^{-1}\vec{c}=\xi\vec{c}$
and this technique converges to the energy closest to $E_0$.
It gives very good results with a well-balanced mixture of IRA and GMRES.
IRA is used to quickly obtain the dominant eigenvalue $\xi_0$
and provides the energy $E=\displaystyle E_0+\frac{1}{\xi_0}$.
The real difficulty lies in the generation of the corresponding Krilov subspace.
It is obviously impossible to invert $(A'-E_0)$ since the $F,G,H$ matrices are contained in $A'$.
The step $\vec{x}_{k+1}=(A'-E_0)^{-1}\vec{x}_k$ is then performed by a GMRES resolution of the
equivalent linear system $(A'-E_0)\vec{x}_{k+1}=\vec{x}_k$. This technique presents considerable interest
especially for excited states which are easily obtained,
independently of the previous convergence of the less excited ones.

\section{Results}

The results presented in this section have been obtained with the spin-dependent S-wave interaction MT I-III: 
\[ V^{\sigma}(r) =  V_r \frac{\exp(-\mu_r r)}{r} - V_a \frac{\exp(-\mu_a r)}{r} \]
The potential parameters
and the value $\displaystyle \frac{\hbar^2}{m}=41.47$ MeV.fm$^{2}$
are the same than those used e.g. in \cite{SSK92,LA82,KG92} 
and are slightly different from those given in the original version \cite{MT69}.
Despite its bare simplicity, this potential turns out to be very efficient in describing
the bulk of low energy properties in the few-nucleon systems.
We will first examine what we call the S-wave approximation, i.e.
the fact that aside from the zero angular momentum of 
the interaction pair, all the angular momenta variables in expansion (\ref{KPW}) are set equal to zero.
The convergence with respect the $l_y,l_z$ expansion will be examined in a second step.

\subsection{Bound states}

In the 4N-system, the bound states exist only for the S=T=0 channel.
In the S-wave approximation the number of FY components is limited to $N_c=4\;\;(N_K=N_H=2)$
(see Table~\ref{tab_amplis}).
The binding energies and r.m.s. radius for the  ground ($^4$He) and first excited ($^4$He$^*$) states
are given in Table~\ref{tab_bs}. 
The corresponding grids are G$_1$ with $N_g=12$ and G$^*_1$ with $N_g=6$, given in Table~\ref{tab_grids1}.
The estimated accuracy in the binding energies is 0.01~MeV but we notice that much less
expensive calculations can provide a precise result as well; e.g. the grid G$_2$
with $N_g=8$ gives also a binding of 30.30~MeV.
In the ground state we remark a good agreement with the best existing calculations \cite{SSK92,KG92}.

The first excitation which, experimentally, corresponds to a $J^{\pi}=0^+$ resonance 0.40~MeV  
above the p+t threshold \cite{TILLEY92}, appears in this model as a loosely bound state.
The binding energy with respect to the N+NNN threshold (E=-8.535~MeV in this model) is 0.257~MeV.
A similar result was found in \cite{F89} in which different versions of the one term separable Yamaguchy 
potential gives a binding energy varying from 0.07 to 0.40~MeV,
depending on their different D-state contribution.
This $0^+$ first excitation has been widely considered in the literature
as being a breathing-mode \cite{BLOM67,WONG71,CVS88}.
Our conclusion is however different. We have calculated the regularized
two-body correlation functions defined by
\begin{equation}\label{CORRELFUNCT}
C_{\alpha_x}(x)= \sum_{\alpha' \, (\alpha'_x=\alpha_x)} \int\!\!\!\!\int dy dz
\cdot \left| \Psi_{\alpha'}(x,y,z) \right|^2
\end{equation}
where $\Psi_{\alpha'}(x,y,z)$ represents the total wave function
component in the $\alpha'$ quantum numbers, and
where $\alpha_x$ denotes the subset of quantum numbers $\alpha$ relative to
the $x$ variable ($l_x,\sigma_x,j_x,\tau_x$).
The summation in (\ref{CORRELFUNCT}) is performed onto one of the two basis, K or H.
Once the total wavefunction is normalized, the correlation functions are normalized according to:
\begin{equation}
\sum_{\alpha_x} \int dx \cdot C_{\alpha_x}(x) =1
\end{equation}

\begin{figure}[hbtp]
\begin{center}\epsfxsize=14cm\mbox{\epsffile{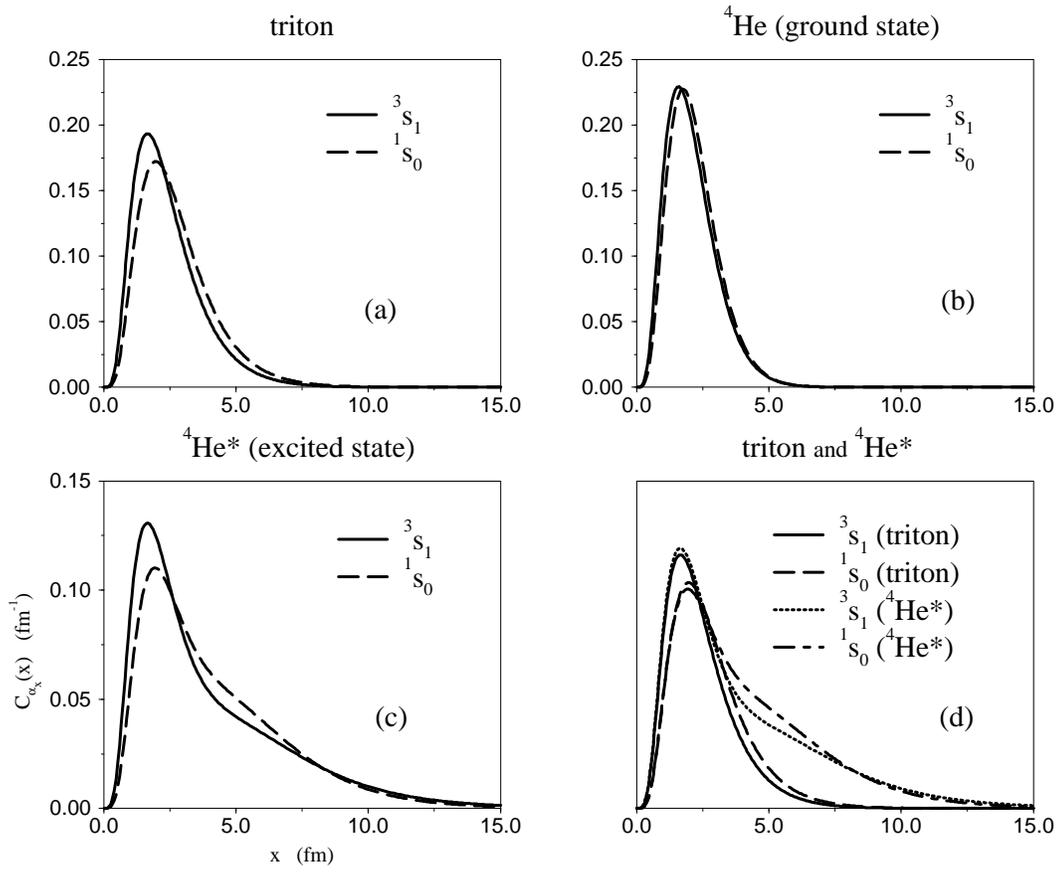}}\end{center}
\caption{Two-body correlation functions $C_{\alpha_x}(x)$ for (a) triton, (b) $^4$He ground and (c) first excited states.
Solid (dashed) line denotes the triplet (singlet) contributions. In (d), the results of $^4$He
first excited state are compared to the triton correlation function suitably scaled.
}\label{figcorrel}
\end{figure}

The results are displayed in Figure~\ref{figcorrel}.
The separated contributions from the singlet and triplet state
are plotted for (a) triton (b) $^4$He ground state and (c) $^4$He first excited state.
The difference between the correlation function for the ground and excited states is
remarkable, both in the shape and in the separated singlet-triplet contributions.
For the excited state one can distinguish the superposition
of two structures with two different length scales,
the short distance part being similar to the triton one.
This suggests a 1+3 structure for the $^4$He excited state, as
can be more clearly seen in plot (d) where the results of (c) are compared
with those of the triton suitably normalized.  
Contrary to what would happen in a breathing mode, the short distance behavior 
of the nucleons is that of an unperturbed triton with the fourth nucleon being simply a remote spectator.

By modifying the MT I-III potential strength we failed to pull the state out 
of the bound region, the 3-body threshold moving in the same direction.
It seems very difficult for a pure strong interaction to generate a first excitation in the continuum.
The right position of this resonance is however a crucial point in any attempt
to describe the low energy data (e.g. p+t) \cite{Fred97}. 
The effect of Coulomb interactions could be enough.
However the inclusion of an ad-hoc repulsive 4-body term 
$V(\rho)=V_0 e^{-\rho^2}$ can also achieve the same result.

The preceding results are only slightly
modified by the inclusion of higher partial waves in the FY expansion.
The effects of these contributions can be seen in Table~\ref{tab_bs2}. Their smallness shows
the validity of the S-wave approximation.
These results have been obtained using the grid G$_2$ with $N_g=8$ for the ground state
and G$^*_1$ with $N_g=6$ for the first excitation for which the corresponding
triton binding energy is $B_3=8.513$~MeV.

\subsection{Elastic N+NNN scattering}

A crucial point in our method is to ensure the proper description of the 3N asymptotic state.
This is used to fix the grid parameters for the $x,y$ variables.
In order to exhibit the stability of our results, we will compare the phase-shifts obtained with several tritons
corresponding to increasing numerical accuracies.
The considered grids, detailed in Table~\ref{grids2}, are T$_4$ with $N_g=6$ and a binding energy $B=8.593$~MeV, 
T$_8$ with $N_g=8$ and $B=8.527$~MeV, and T$_{10}$ with $N_g=8$ and $B=8.535$~MeV.
We recall that the precise value for the 3N binding energy is $B=8.535$~MeV.
The grid parameters for the z-variable depends substantially on the relative kinetic energy.
A zero energy calculation requires a relatively large value of $z_{N}$ but very few points inside 
are sufficient to describe an asymptotic linear behavior. On the contrary as far as the energy increases,
the value of $z_{N}$ can be decreased but the oscillations in the relative
wave functions demand an increasingly big number of points.
A typical grid for the case $E=0$ is   G$_z\equiv\{ 10 , 1.20 , 19.0 ; 03 , 1.00 , 34.0\}$  

We have shown in Table~\ref{DEPHAN_NNN} the phase-shifts for different $(S,T)$
channels as a function of the center of mass energy.
For all of them, we have arbitrarily chosen, as in ref.~\cite{TJON76}, the determination $\delta(E=0)=180^{\circ}$. 
For the S=0 case the comparison between grids T$_4$ and T$_8$ has been made, showing a good stability
despite the fact that grid T$_4$ gives only a poor description of the asymptotic state.
As it has been already emphasized in section 2, the key point in our approach
to the scattering problem lies in the coherence between 
the asymptotic state and the numerical solution of the 4N problem, rather than in a very precise description of it. 
The results corresponding to grid T$_8$ are considered as converged.

A zero energy calculation directly provides us with the scattering length.
The results, given in Table~\ref{tab_sl}, show a high stability
with respect to grid variations (T$_8$ and T$_{10}$) and our estimated accuracy is given in column 3.
These values are in agreement with the existing published
calculations \cite{YAKOVLEV95}.
For $T=1$ they are close to those obtained in \cite{TJON76} although with the original potential parameters.

The remaining low energy parameters have been extracted from the phase-shifts and are given in Table~\ref{tab_lep}
for the different $(S,T)$ channels .
They are defined in the effective range expansion:
\begin{equation}\label{MERA}
g_0(q)= q\cdot\cot\delta(q)=\chi(q)\left[-\frac{1}{a} +\frac{1}{2} \cdot r_0 \cdot q^2+v_0 \cdot q^4\right]
\end{equation}
where $\chi(q)=1$ in the usual case but has to be modified in the presence of a near threshold singularity. 
According to \cite{ADHIK81}, we have taken the form ($q_0$ real): 
\begin{equation}\label{pole}
 \chi(q)=\frac{1}{1-\left(\frac{q}{q_0}\right)^2} 
\end{equation}

\begin{figure}[hbtp]
\begin{center}
\epsfxsize=14.0cm\epsfysize=14.cm\mbox{\epsffile{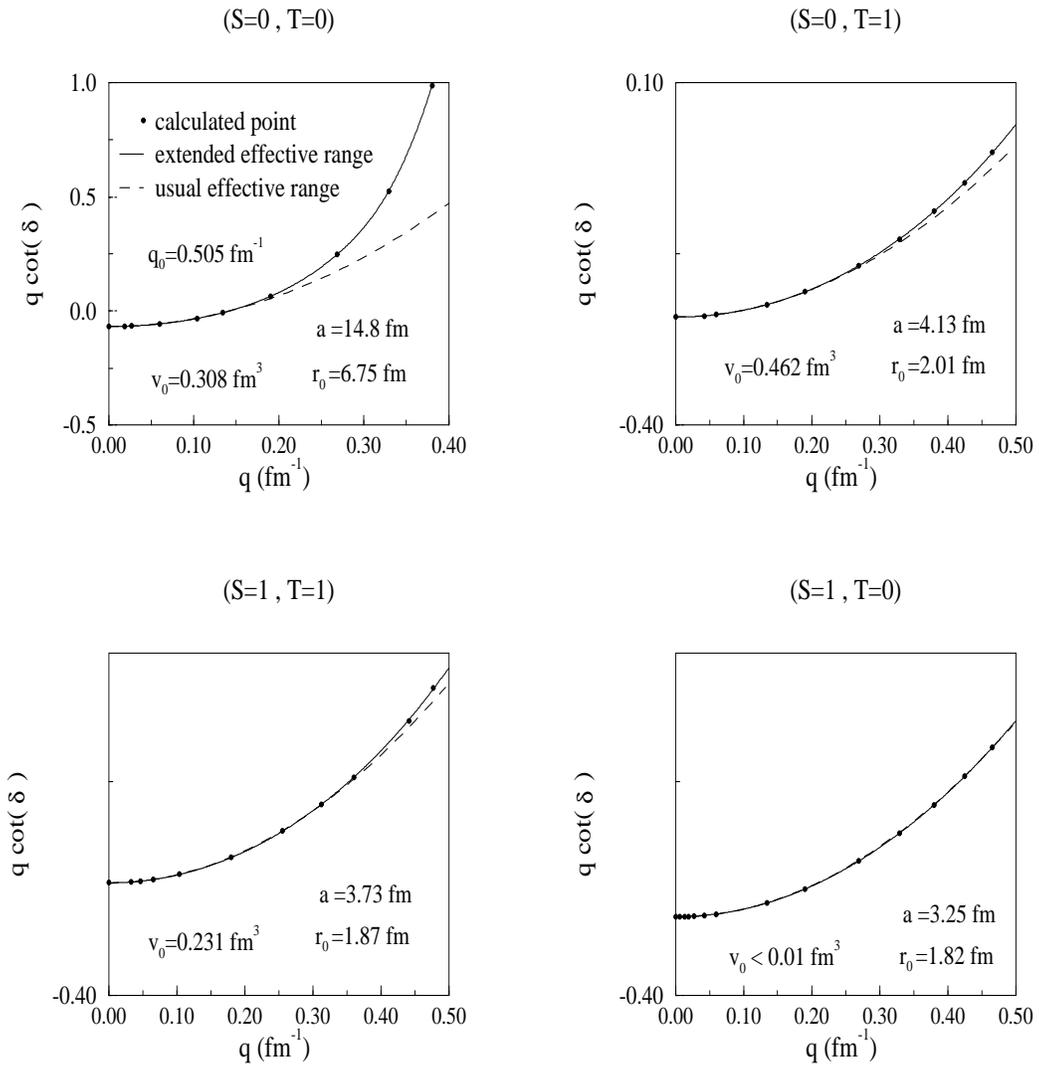}}
\end{center}
\caption{Effective range expansion, $q$ being the center of mass momentum.
The usual one (dashed curve) gives
an quite accurate description of the scattering amplitude, except in the $(S=0,T=0)$ case.
The full expansion (\protect \ref{MERA}) (solid line) provides a perfect fit of
the calculated points.}\label{fig_kcotd}
\end{figure}

The validity of the expansion (\ref{MERA}) in the energy region below the first inelastic threshold 
is displayed in Figure~\ref{fig_kcotd}.
The usual effective range approximation, i.e. $\chi(q)=1$ and
the $q^4$ contribution neglected in (\ref{MERA}), works very well
in the whole energy domain for all but the $(S=0,T=0)$ channel,
in which the existence of a near threshold $^4$He excited state requires
the explicit inclusion of the pole contribution (\ref{pole}). 
It is worth noticing that the contributions coming from the $v_0$ term
are very small and have been included only for completeness.
On the contrary the pole contribution, existing only in the $(S=0,T=0)$ channel, is essential.
Expansion (\ref{MERA}) provides a very accurate parametrization of the
S-wave scattering amplitude in all the energy region below the break-up threshold.

We would like to emphasize here the coherence between the $^4$He$^*$
and the scattering results in the $(S=0,T=0)$ channel.
Inserting the effective range expansion (\ref{MERA}) in the S-wave scattering amplitude
\[  f_0(q)={1\over g_0(q)-iq}\]
produces an imaginary pole in the upper complex q-plane with value $q^*=0.095i$. The corresponding energy 
$\displaystyle E^*={2\over3}{\hbar^2 {q^*}^2\over m}=0.25$~MeV is in close agreement
with the direct calculation given in Table~\ref{tab_bs}.

A step beyond the S-wave approximation has been taken by keeping the S-wave interaction alone 
but allowing non zero values in the angular momentum expansion of the FY amplitudes.
Table~\ref{ldifl} shows the sensibility of the scattering lengths values
when the maxima of $l_y$ and $l_z$ vary from 0 to 2.
In the particular interaction model we are considering,
the number of channels describing the 3N asymptotic state remains 
the same whereas the number of channels $N_c$ of the 4N problem considerably increases.
These results have been  obtained with the grid T$_8$ and $N_g=8$ completed with a suitable z-grid .
The values are well converged with $l_y,l_z=0,1$ except in the $(S=0,T=0)$ case,
where the big value of the scattering length makes
this state rather sensitive to small parameter variations.

\subsubsection*{The n+t cross section}

\begin{figure}[htbp]
\begin{center}
\epsfxsize=14.0cm\mbox{\epsffile{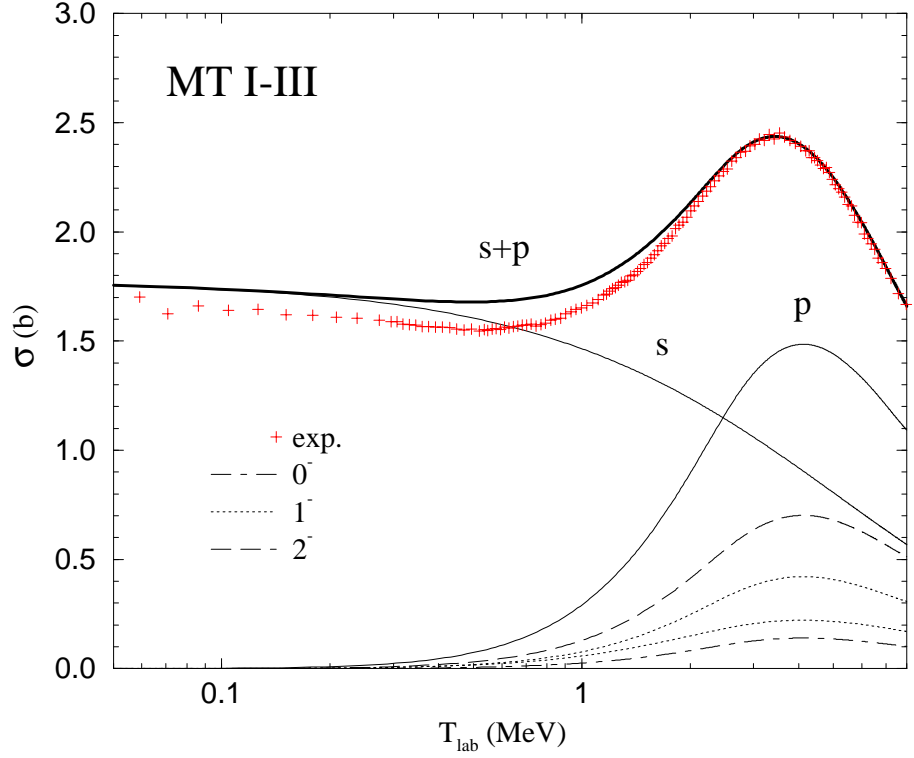}}
\end{center}
\caption{The n+t cross section $\sigma$ calculated with MT~I-III potential
is compared to experimental data.
The s-wave contribution (s, solid line) is slightly overestimated due to the overestimated scattering lengths.
The p-wave contribution (p, solid line) dominates in the resonance region and is responsible for
the very nice agreement (s+p thick line) with the experimental total cross section (+'s).
The $L=1,S=1$ contribution is split by statistical factors into its $J^{\pi}=0^-,1^-,2^-$
components (dot-dashed, dotted, dashed curves),
whereas the $L=1,S=0$ one corresponds to a pure $J^{\pi}=1^-$ (dotted curve) partial wave.}
\label{nt1}
\end{figure}

Of particular interest is to apply the preceding results to the description of n+t cross section.
On the one hand this is a pure $T=1$ channel, free from the difficulties related to the Coulomb interactions.
On the other hand accurate low energy scattering data exist \cite{PHILL80},
showing a structure at neutron laboratory energy 
$T_{lab}\approx 4$ MeV which is supposed to be created by a series
of P-waves resonances \cite{TILLEY92}.
The calculations discussed in the preceding section have thus been completed
up to the first 3-body break-up threshold by the inclusion
of the first negative parity states $J^{\pi}=0^-,1^-,2^-$ corresponding to n+t relative P-waves.
The resulting total cross sections are plotted in Figure~\ref{nt1}.
The contributions from n+t relative S- and P-waves are distinguished. 
We notice that the MT I-III model conserves separately L and S and consequently
the $J^{\pi}=0^-,1^-,2^-$ states coming from $L=1,S=1$ are degenerate.
The corresponding cross sections differ only by statistical factors.
The remaining  $J^{\pi}=1^-$ state comes from an $L=1,S=0$ coupling.
In view of these results, several remarks are in order:
\begin{enumerate}
\item The scattering lengths obtained in the S-wave approximation (Table~\ref{tab_sl})
gives a slightly overestimated value for the zero energy cross section $\sigma(0)=1.85$~b.
The experimental extrapolated zero energy cross section is
$\sigma(0)=1.70\pm 0.03$~b \cite{SEAG80,PHILL80}.
However the inclusion of higher partial waves in the FY expansion (Table~\ref{ldifl}) significantly 
reduces the S-wave approximation result to $\sigma(0)=1.77$~b in closer agreement with experiment. 

\item The comparison of the separated spin contribution 
is not possible since the values of the spin-dependent scattering lengths,
summarized in Table~\ref{tab3}, are still controversial.

\item Despite the simplicity of this model, the agreement with experimental data in the resonance region, is very good.
The n+t P-waves resonances are generated by an NN S-wave interaction alone.
An effective 1+3 P-wave potential is created due to the exchange mechanism between the four nucleon.
We remark however that a first attempt to describe this cross section with $l_z=1$ in the K components but
keeping zero all the remaining angular momenta in (\ref{KPW}) failed. The
incoming neutron seems thus to be more sensible to the virtual excitations of the triton
than to the NN pair interactions themselves.

\item We have calculated the contribution coming from the n+t relative D-waves.
They are given by the positive parity states $L=2$, $S=0,1$. The corresponding phase-shifts at 6~MeV c.m.
kinetic energy
are $\delta_{S=0}=-3.3^{\circ}$ and $\delta_{S=1}=-2.4^{\circ}$, which contributes only a few mb
to the total cross section.
The results displayed Figure \ref{nt1} can so be considered as fully converged in the MT~I-III model.
\end{enumerate}

\subsection{First inelastic channel}

The last point to be presented in this work concerns the reaction
\begin{eqnarray*}
\mbox{p+$^3$H} &\rightarrow& \mbox{p+$^3$H}  \cr
      &\rightarrow& \mbox{n+$^3$He}  \cr
      &\rightarrow& \mbox{d+d}    
\end{eqnarray*}
In the isospin approximation employed throughout this work, both n+$^3$He and p+t  thresholds are degenerate
and correspond to the N+NNN of the preceding section.
The Pauli principle imposes for the deuteron-deuteron channel $(-)^{L+S+T}=+1$.
If we assume the final state d+d to be in a relative S-wave
one has $L=T=0$ and  $J^\pi=0^+,2^+$.
The $J^\pi=2^+$ state requires a relative angular momentum $l_z=2$ in the initial N+NNN channel
and it is expected to give small contribution at very low d+d energy.
We will consider only the $(J^\pi=0^+, T=0)$ state.

The S-matrix is defined according to (\ref{smatrix}).
We present in Table~\ref{Smat} the S-matrix elements at several energies,
chosen with respect to the NN+NN threshold. N+NNN is referred to as channel~1, NN+NN as channel~2.
The symmetry and unitarity properties are there fulfilled at the level of $10^{-3}$,
what corresponds to the accuracy of our results.
 
\begin{figure}[hbtp]
\begin{center}\epsfxsize=14cm\mbox{\epsffile{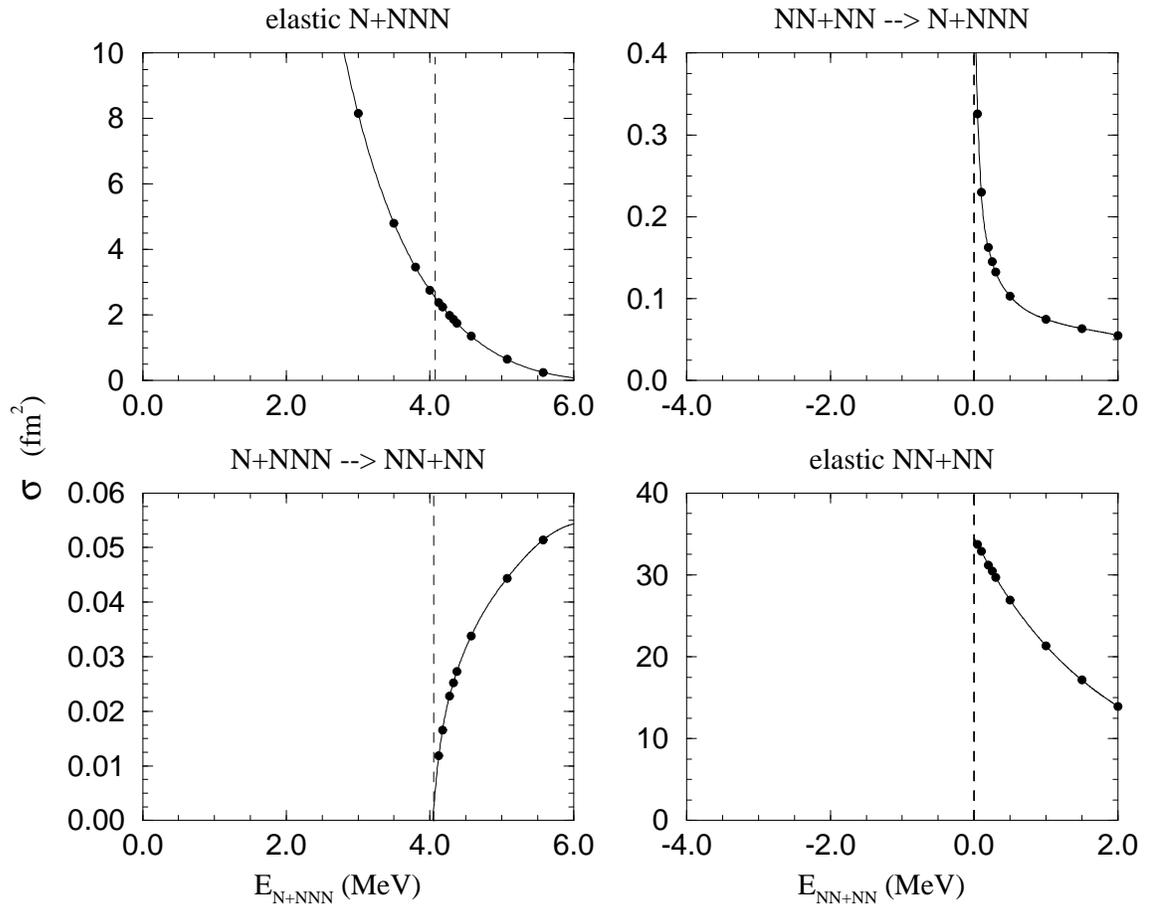}}\end{center}
\caption{Elastic and inelastic cross sections (solid curves), in fm$^2$, for the
coupled channels N+NNN-NN+NN. They are interpolated between the calculated values (filled circles).
The energies are given in the center of mass of the incident channel.}
\label{SIGINEL}
\end{figure}

With the conventions used above, the scattering cross sections in a given partial wave ($J^{\pi}$)
in presence of several channels $a,a',\ldots$ are given by:
\begin{eqnarray}
\sigma^{(J)}_{a \rightarrow a}&=&\frac{\pi}{q_{a}^2} \cdot
\frac{\widehat{J}}{\widehat{S_{a_1}} \cdot \widehat{S_{a_2}}} \cdot \left|1-S_{a a} \right|^2
\\
\sigma^{(J)}_{a \rightarrow a'}&=&\frac{\pi}{q_{a}^2} \cdot
\frac{\widehat{J}}{\widehat{S_{a_1}} \cdot \widehat{S_{a_2}}} \cdot \left| S_{a' a} \right|^2
\end{eqnarray}
where the notation $\hat{J}$ holds for $2J+1$. $S_{a_i}$ denotes the spin of the colliding cluster $a_i$.
The corresponding values with $a,a'\equiv$~N+NNN,NN+NN are given in Figure~\ref{SIGINEL} (filled circles).
We remark no accident in the N+NNN cross section when the inelastic NN+NN threshold is open.
The NN+NN~$\rightarrow$~N+NNN cross section displays the usual $1/v$ law of the inelastic process
and the crossed channel N+NNN~$\rightarrow$~NN+NN the expected $\sqrt{E}$ law.
The elastic d+d cross section has been calculated neglecting Coulomb interactions
which will make this quantity divergent.
However, due to the absence of nearthreshold singularity, one
can expect small corrections to the low energy parameters obtained in this way,
like in the n+n versus p+p case.

\begin{figure}[hbtp]
\begin{center}\epsfxsize=10cm\mbox{\epsffile{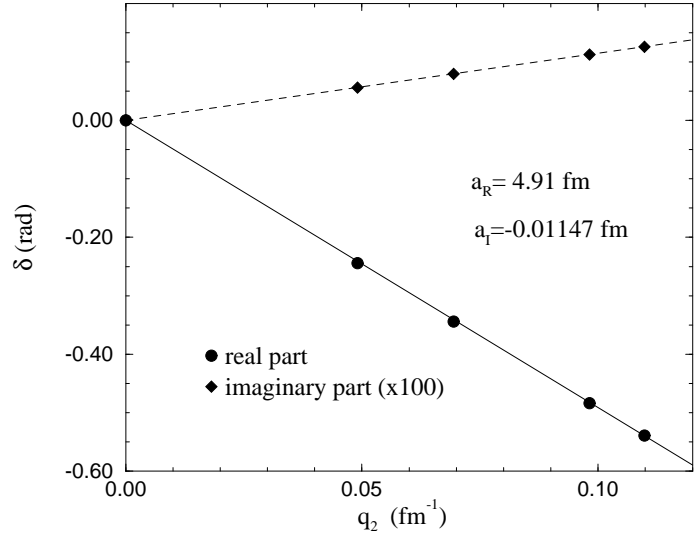}}\end{center}
\caption{Low energy behavior of the deuteron-deuteron phase-shift $\delta$ and
determination of the scattering length. Its real or imaginary parts $a_R$ or $a_I$ are
deduced from the proportionality between the real (solid curve)
or imaginary (dashed curve) parts of
the phase shift and the center of mass momentum $q_2$.}\label{ddd1}
\end{figure}

Of particular interest is the extraction of the imaginary part of the strong d+d scattering length,
a quantity which controls the fusion rate in the process d+d~$\rightarrow$~n+$^3$He.
This can be done either from the linear behavior of the d+d phase-shifts $\delta_{22}=-(a_R-ia_I)q_2$ 
(with $\displaystyle S_{22}=e^{2i\delta_{22}}$), as displayed in Fig.~\ref{ddd1},
or from the inelasticity in the non diagonal S-matrix element which behaves like $|S_{12}|^2=-4a_Iq_2\;(1+2a_Iq_2)$.
Both methods agree with high accuracy and lead to the following values:
\begin{eqnarray}
a_R&=&+4.91\pm0.02 \;\; \mbox{fm}\\
a_I&=&-0.0115\pm0.0001 \;\; \mbox{fm}
\end{eqnarray}
We remark a very small value of $a_I$ that should be only slightly modified
once the Coulomb interaction is switched on.
This small value is due to the small overlapping between the K and H configurations which
respectively govern the N+NNN and NN+NN asymptotic states.

\section{Conclusion}

We have presented the first solution of Faddeev-Yakubovsky equations
in configuration space for the scattering states in the 4N system.
They concern both the N+NNN elastic scattering and its coupling to the first inelastic NN+NN channel.

The results presented here have been obtained with an S-wave model interaction
and in the isospin symmetry approximation, i.e. neglecting the Coulomb and mass difference effects.

The N+NNN elastic phase-shifts have been calculated for the different spin and isospin channels. 
The low energy parameters have been extracted and the validity
of the effective range approximation in the energy region below the 3-body break-up threshold has been analyzed.
We have in particular found the importance of including the pole contribution
in the $(S=0,T=0)$ case due to the vicinity of the first $J^{\pi}=0^+$ excitation.
We have also found that in the framework of our model this $0^+$ state
is bound at 0.25~MeV below the 3N threshold.
The study of the two-body correlation functions showed that the structure of this state is a 1+3 configuration
rather than a breathing mode, as it is usually accepted.
The coherence between the binding energy of this state and the scattering results has been emphasized.

The n+t scattering cross section has been treated with a special interest
and the first negative parity states have been included, to account for the structure experimentally observed.
The elastic cross section is well described by the simple MT~I-III interaction, especially in the resonance region.
The n+t P-waves resonance is thus reproduced by a NN pure S-wave interaction.
This shows the difficulty to understand this structure in terms of the NN interaction alone.
It is created by the direct and exchange mechanism between the incoming neutron and the target nucleons.

The S-matrix coupling the N+NNN and NN+NN channels has been obtained
as well as the corresponding cross sections.
The analysis of these results allows the extraction of the NN+NN scattering length, whose imaginary part
controls the fusion process d+d~$\rightarrow$~n+$^3$He.
Its value turned out to be very small ($a_I\simeq 0.011$ fm).

Further applications of this formalism including Coulomb interactions
and more realistic potentials are in process.

\bigskip
{\bf Acknowledgements:}
We are indebted to Pr.~C.~Gignoux for his continuous teaching and helpful discussions during the
elaboration of this work. 
The numerical calculations were initiated on the Cray-T3D/T3E at the CGCV (Centre Grenoblois de Calcul Vectoriel,
CEA), and completed on the Cray-T3E at the IDRIS (Institut du D\'eveloppement et
des Ressources en Informatique Scientifique, CNRS).
We are grateful to the staff members of these two organizations for their kind hospitality and useful advice.

\bigskip
\appendix

\section*{ }

This appendix aims at the complete expression of the functions $f,g,h$ appearing in (\ref{fse}) in terms of
the corresponding quantum numbers $\alpha$ and $\alpha'$.
By projection onto the appropriate K and H basis, the r.h.s. of the equations (11) gives
raise to integral terms as shown in (13), involving some very complicated
functions $f$, $h$, $g$, generated by permutations operators. 

\subsection{Permutation operators}
The permutation operators $P_{ij}$ are completely defined by their action upon each ket
of a complete basis. This basis is chosen to be an ordered quadruplet $|q_1 q_2 q_3 q_4>$,
where the given $i^{th}$ value represents the state of
the $i^{th}$ particle, including the space, spin and isospin degrees of freedom.
This basis is a generalization of the one given in (\ref{baser}).
It is assumed that $P_{ij}$ corresponds to an exchange
between the $i^{th}$ and $j^{th}$ set of quantum numbers, e.g.
$$
P_{34}|q_1 q_2 q_3 q_4> =|q_1 q_2 q_4 q_3>
$$
More complex permutations are then transparent, e.g.
$$
P_{23}P_{34}|q_1 q_2 q_3 q_4> =|q_1 q_4 q_2 q_3>
$$  
The way the permutation operators act upon the coupled bases (K or H) can be
deduced from this basic feature. In the configuration space, e.g., a single operator
results in a generalized rotation of the Jacobi coordinates. 
Note that because of the symmetry of the bases, the contribution of 
the operators $P_{13}$, $P_{13}P_{34}$ and $P_{14}P_{23}$ are identical with those
of $P_{23}$, $P_{23}P_{34}$ and $P_{13}P_{24}$. 

\subsection{${\cal{H}}$ functions} 
All the permutation operations can be seen as, at most, two successive rotations, each of
them involving only two coordinates, e.g. first of all $y$ and $z$, then $x$ and one of the preceding
rotated coordinates.
So it is convenient to use the functions appearing in the 3-body problem \cite{ISN7211},
defined as follows.

Suppose we are to calculate the projection 
of a given expression $[Y_{l_x'}(\hat{x}') Y_{l_y'}(\hat{y}')]_L F(x',y')$,
$F$ being an arbitrary function of $x'$ et $y'$,
onto a bipolar-harmonic basis $[Y_{l_x}(\hat{x}) Y_{l_y}(\hat{y})]_L$, where the following relation holds:
\begin{equation}\label{MATROTXY}
\left( \begin{array}{c} \vec{x}' \\ \vec{y}' \end{array} \right) = \left( \begin{array}{cc} a&b\\c&d \end{array} \right)
\left( \begin{array}{c} \vec{x} \\ \vec{y} \end{array} \right)
\end{equation}
We define ${\cal{H}}$ functions such that:
$$
\displaystyle  \int \!\!\!\! \int d\hat{x} d\hat{y} \cdot  [Y_{l_x}(\hat{x}) Y_{l_y}(\hat{y})]^*_L 
 \cdot [Y_{l_x'}(\hat{x}') Y_{l_y'}(\hat{y}')]_L \cdot  F(x',y') =
\displaystyle \frac{1}{2} \int_{-1}^1 du \cdot  {\cal{H}}^L_{l_x l_y , l_x' l_y'}(x,y,u) 
 \cdot F(x',y')
$$
where, in the integral term, $x'$ et $y'$ are obtained from
(\ref{MATROTXY}) with the constraint $\cos(\widehat{\hat{x},\hat{y}})=u$.
These functions are given for example by:
\begin{eqnarray*}
 {\cal{H}}^L_{l \lambda, l' \lambda'}(x,y,u)&=&
\displaystyle \sum_{k,l_1,l_2,\lambda_1,\lambda_2,l_0,\lambda_0
(l_1+l_2=l',\lambda_1+\lambda_2=\lambda')}
(-)^{l'+\lambda'+L+k} \cdot a^{l_1} b^{l_2} c^{\lambda_1} d^{\lambda_2}
\hat{l'} \hat{\lambda'} \hat{k} \hat{l_0} \hat{\lambda_0}\\
&\cdot&\sqrt{\frac{(2l')!(2\lambda')!\;\hat{l}\hat{\lambda}}
{(2l_1)!(2l_2)!(2\lambda_1)!(2\lambda_2)!}}
 \troisj{k}{l_0}{l}{0}{0}{0}  \troisj{k}{\lambda_0}{\lambda}{0}{0}{0}
  \troisj{l_1}{\lambda_1}{l_0}{0}{0}{0} \\
&\cdot&\displaystyle \troisj{l_2}{\lambda_2}{\lambda_0}{0}{0}{0}
 \sixj{\lambda}{l}{L}{l_0}{\lambda_0}{k}
 \neufj{l_1}{l_2}{l'}{\lambda_1}{\lambda_2}{\lambda'}{l_0}{\lambda_0}{L} 
 \cdot \frac{x^{l_1+\lambda_1}y^{l_2+\lambda_2}}{(x')^{l'} (y')^{\lambda'}}
 \cdot  P_k(u)
\end{eqnarray*}
where $\hat{l}=2l+1$ et $P_k$ stands for the $k^{th}$-Legendre polynomial.

Obviously, in the four-body case, one has to deal with cumbersome recoupling,
to isolate bipolar harmonics from the more complex coupling scheme (\ref{devcoupling}).
Nevertheless all the work can be built upon this basic functions.

\subsection{The only non-zero $f$ function}
$f_{\alpha,\alpha'}$ is non-zero when both $\alpha$ and $\alpha'$ are of H type.
Let us define then $x^f_{\alpha \alpha'}$, $y^f_{\alpha \alpha'}$,
 $z^f_{\alpha \alpha'}$ by
$$
\left( \begin{array}{c} \vec{x}^f_{\alpha \alpha'} \\ \vec{y}^f_{\alpha \alpha'} \\
\vec{z}^f_{\alpha \alpha'} \end{array} \right) = 
\left( \begin{array}{ccc}
0 & 1 & 0\\
1 & 0 & 0\\
0 & 0 & -1
\end{array} \right)
\left( \begin{array}{c} \vec{x}  \\ \vec{y}  \\ \vec{z}  \end{array} \right)
$$
$f_{\alpha,\alpha'}$ is given by:
$$
f_{\alpha,\alpha'}=
t_{\alpha,\alpha'}
 \cdot \displaystyle \delta_{l_z,l_z'} \delta_{\sigma_x',\sigma_y} 
 \delta_{\sigma_x,\sigma_y'} \delta_{l_x,l_y'} \delta_{l_x',l_y} 
 \delta_{j_x,j_y'}  \delta_{j_x',j_y} \delta_{j_{xy},j_{xy}'}
 \cdot (-)^{j_x+j_y-j_{xy}+l_z}
$$
where the isospin contribution $t_{\alpha,\alpha'}$ is 
$t_{\alpha,\alpha'}  = 
\delta_{\tau_x,\tau_y'} \delta_{\tau_x',\tau_y} \cdot  (-)^{\tau_x+\tau_y-T}$.

\subsection{$h$ functions}
There are two cases where the $h$ functions can be non-zero: when the amplitude $\alpha'$
is of K type.

\subsubsection{$\alpha$ of K type}
$x^h_{\alpha \alpha'}$, $y^h_{\alpha \alpha'}$,
 $z^h_{\alpha \alpha'}$ are then defined by
$\left( \begin{array}{c} \vec{x}^h_{\alpha \alpha'} \\ \vec{y}^h_{\alpha \alpha'} \\ \vec{z}^h_{\alpha \alpha'} \end{array} \right) = 
\left( \begin{array}{ccc}
\frac{1}{2} &  \frac{\sqrt{3}}{2} & 0\\
 \frac{\sqrt{3}}{2} &  -\frac{1}{2} & 0\\
0&0&1
\end{array} \right)
\left( \begin{array}{c} \vec{x}  \\ \vec{y}  \\ \vec{z}  \end{array} \right)$
and the complete expression for $h_{\alpha \alpha'}$ is:
\begin{eqnarray*}
h_{\alpha,\alpha'}(x,y,z,u)&=&\varepsilon \cdot
t_{\alpha,\alpha'}
 \cdot \displaystyle \sum_{l_{xy},\sigma} \delta_{l_z,l_z'} \delta_{j_z,j_z'}  \delta_{J_3,J_3'}
 \cdot \aneufj{l_x}{\sigma_x}{j_x}{l_y}{s_3}{j_y}{l_{xy}}{\sigma}{J_3}
 \cdot \aneufj{l_x'}{\sigma_x'}{j_x'}{l_y'}{s_2}{j_y'}{l_{xy}}{\sigma}{J_3}\\ 
&\cdot&\displaystyle (-)^{s_1+s_2-\sigma_x+s_2+\sigma_x'-\sigma}
 \cdot \asixj{s_2}{s_1}{\sigma_x}{s_3}{\sigma}{\sigma_x'}
 \cdot \frac{x y z}{x^h_{\alpha \alpha'} y^h_{\alpha \alpha'} z^h_{\alpha \alpha'}} \\
&\cdot&\displaystyle {\cal{H}}^{l_{xy}}_{l_x l_y , l_x' l_y'}(x,y,u) 
\end{eqnarray*}
where $\displaystyle \aneufj{j_1}{j_2}{j_3}{j_4}{j_5}{j_6}{j_7}{j_8}{j_9}=
\sqrt{\hat{j_3}\hat{j_6}\hat{j_7}\hat{j_8}} 
 \cdot \neufj{j_1}{j_2}{j_3}{j_4}{j_5}{j_6}{j_7}{j_8}{j_9}$, $\hat{j}=
2j+1$, 
$\displaystyle t_{\alpha,\alpha'}  = (-)^{t_1+t_2-\tau_x+t_2+\tau_x'-T_3'}
\cdot \delta_{T_3,T_3'} \cdot \asixj{t_2}{t_1}{\tau_x}{t_3}{T_3}{\tau_x'}$
and with the constraint $\cos(\widehat{\hat{x},\hat{y}})=u$.

\subsubsection{$\alpha$ of H type}
$x^h_{\alpha \alpha'}$, $y^h_{\alpha \alpha'}$,
 $z^h_{\alpha \alpha'}$ are now defined by
$\left( \begin{array}{c} \vec{x}^h_{\alpha \alpha'} \\ \vec{y}^h_{\alpha \alpha'} \\
\vec{z}^h_{\alpha \alpha'} \end{array} \right) = 
\left( \begin{array}{ccc}
 -\frac{1}{\sqrt{3}} & 0 & -\frac{\sqrt{2}}{\sqrt{3}}\\
0&1&0\\
 \frac{\sqrt{2}}{\sqrt{3}} & 0 &  -\frac{1}{\sqrt{3}}
\end{array} \right)
\left( \begin{array}{c} \vec{x}  \\ \vec{y}  \\ \vec{z}  \end{array} \right)$
and the corresponding expression for $h_{\alpha \alpha'}$ is:
\begin{eqnarray*}
h_{\alpha,\alpha'}(x,y,z,u)&=&\varepsilon \cdot
t_{\alpha,\alpha'}
 \cdot \displaystyle \sum_{l_{yz}',J_2'} \delta_{l_x',l_y} \delta_{\sigma_x',\sigma_y}
 \cdot (-)^{j_{xy}+l_z-J+j_y+J_2'-J+l_x+l_z-l_{yz}'}
 \cdot \asixj{l_z}{j_x}{J_2'}{j_y}{J}{j_{xy}}\\ 
&\cdot&\displaystyle  \asixj{j_x'}{j_y'}{J_3'}{j_z'}{J}{J_2'}
 \cdot \aneufj{l_y'}{s_1}{j_y'}{l_z'}{s_2}{j_z'}{l_{yz}'}{\sigma_x}{J_2'}
 \cdot \asixj{l_z}{l_x}{l_{yz}'}{\sigma_x}{J_2'}{j_x}\\ 
&\cdot&\displaystyle \frac{x y z}{x^h_{\alpha \alpha'} y^h_{\alpha \alpha'} z^h_{\alpha \alpha'}} 
 \cdot {\cal{H}}^{l_{yz}'}_{l_x l_z , l_y' l_z'}(x,z,u) 
\end{eqnarray*}
with $t_{\alpha,\alpha'}  = 
\delta_{\tau_x',\tau_y} \cdot  (-)^{\tau_x+\tau_y-T}
\cdot \asixj{\tau_x'}{t_3}{T_3'}{t_4}{T}{\tau_x}$
and the constraint $\cos(\widehat{\hat{x},\hat{z}})=u$.

\subsection{$g$ functions}
There are two cases where the $g$ functions can be non-zero: the amplitude $\alpha$
must be of K type.

\subsubsection{$\alpha'$ of K type}
It is necessary to define in this case $x^g_{\alpha \alpha'}$, $y^g_{\alpha \alpha'}$,
 $z^g_{\alpha \alpha'}$ and an intermediate coordinate $\vec{y}_0$ such that:
$$
\left( \begin{array}{c} \vec{x}^g_{\alpha \alpha'} \\ \vec{y}^g_{\alpha \alpha'} \\ \vec{z}^g_{\alpha \alpha'} \end{array} \right) = 
\left( \begin{array}{ccc}
1&0&0\\
 0& \frac{1}{3} & \frac{2\sqrt{2}}{3} \\
 0& \frac{2\sqrt{2}}{3} &  -\frac{1}{3} 
\end{array} \right)
\left( \begin{array}{c} \vec{x}^g_{\alpha \alpha'} \\ \vec{y}_0  \\ \vec{z}  \end{array} \right) 
\; \mbox{and} \;
\left( \begin{array}{c} \vec{x}^g_{\alpha \alpha'} \\ \vec{y}_0 \\ \vec{z} \end{array} \right) = 
\left( \begin{array}{ccc}
 \frac{1}{2} & \frac{\sqrt{3}}{2} & 0\\
 \frac{\sqrt{3}}{2} &  -\frac{1}{2} & 0\\
0&0&1
\end{array}
\right)
\left( \begin{array}{c} \vec{x}  \\ \vec{y}  \\ \vec{z}  \end{array} \right)
$$ 
The corresponding expression for $g_{\alpha \alpha'}$ is:
\begin{eqnarray*}
g_{\alpha,\alpha'}(x,y,z,u,v)&=&
\displaystyle \frac{1}{2} \cdot  t_{\alpha,\alpha'}
 \cdot \sum_{l_{xy},\sigma,l_{xy}',\sigma',l'_{yz},L,S,\lambda}
\aneufj{l_x}{\sigma_x}{j_x}{l_y}{s_3}{j_y}{l_{xy}}{\sigma}{J_3}
 \cdot \aneufj{l_x'}{\sigma_x'}{j_x'}{l_y'}{s_4}{j_y'}{l_{xy}'}{\sigma'}{J_3'}\\ 
&\cdot&\displaystyle \aneufj{l_{xy}}{\sigma}{J_3}{l_z}{s_4}{j_z}{L}{S}{J}
 \cdot  \aneufj{l_{xy}'}{\sigma'}{J_3'}{l_z'}{s_2}{j_z'}{L}{S}{J}
 \cdot \asixj{l_x'}{l_y'}{l_{xy}'}{l_z'}{L}{l_{yz}'}\\ 
&\cdot&\displaystyle \asixj{l_x'}{\lambda}{l_{xy}}{l_z}{L}{l_{yz}'}
 \cdot (-)^{s_1+s_2-\sigma_x+s_2+\sigma'-S}
 \cdot \asixj{s_2}{s_1}{\sigma_x}{s_3}{\sigma}{\sigma_x'}\\ 
&\cdot&\displaystyle \asixj{s_2}{\sigma_x'}{\sigma}{s_4}{S}{\sigma'}
 \cdot \frac{x y z}{x^g_{\alpha \alpha'} y^g_{\alpha \alpha'} z^g_{\alpha \alpha'}} \\
&\cdot&\displaystyle {\cal{H}}^{l_{xy}}_{l_x l_y , l_x' \lambda}(x,y,u) 
\cdot {\cal{H}}^{l_{yz}'}_{\lambda l_z , l_y' l_z' }(y_0,z,v)
\end{eqnarray*}
with the constraints $\cos(\widehat{\hat{x},\hat{y}})=u$ and $\cos(\widehat{\hat{y}_0,\hat{z}})=v$, and
where:
$$ 
\displaystyle t_{\alpha,\alpha'}  = 
(-)^{t_1+t_2-\tau_x+t_2+T_3'-T}
 \cdot \asixj{t_2}{\tau_x'}{T_3}{t_4}{T}{T_3'}
 \cdot \asixj{t_2}{t_1}{\tau_x}{t_3}{T_3}{\tau_x'}
$$

\subsubsection{$\alpha'$ of H type}
We define again $x^g_{\alpha \alpha'}$, $y^g_{\alpha \alpha'}$,
 $z^g_{\alpha \alpha'}$ and an intermediate coordinate $\vec{y}_0$ in the following way:
$$
\left( \begin{array}{c} \vec{x}^g_{\alpha \alpha'} \\ \vec{y}^g_{\alpha \alpha'} \\ \vec{z}^g_{\alpha \alpha'} \end{array} \right) = 
\left( \begin{array}{ccc}
1&0&0\\
 0& -\frac{1}{\sqrt{3}} & \frac{\sqrt{2}}{\sqrt{3}} \\
 0& \frac{\sqrt{2}}{\sqrt{3}} & \frac{1}{\sqrt{3}} 
\end{array} \right)
\left( \begin{array}{c} \vec{x}^g_{\alpha \alpha'}  \\ \vec{y}_0  \\ \vec{z}  \end{array} \right)
\; \mbox{and} \;
\left( \begin{array}{c} \vec{x}^g_{\alpha \alpha'} \\ \vec{y}_0 \\ \vec{z} \end{array} \right) = 
\left( \begin{array}{ccc}
 \frac{1}{2} &  \frac{\sqrt{3}}{2} & 0\\
 \frac{\sqrt{3}}{2} &  -\frac{1}{2} & 0\\
0&0&1
\end{array} \right)
\left( \begin{array}{c} \vec{x}  \\ \vec{y}  \\ \vec{z}  \end{array} \right)
$$
The corresponding expression for $g_{\alpha \alpha'}$ is now:
\begin{eqnarray*}
g_{\alpha,\alpha'}(x,y,z,u,v)&=&
\displaystyle \frac{\varepsilon}{2}  \cdot  t_{\alpha,\alpha'}
 \cdot \sum_{l_{xy},\sigma,l_{xy}',l'_{yz},L,S,\lambda}
\aneufj{l_x}{\sigma_x}{j_x}{l_y}{s_3}{j_y}{l_{xy}}{\sigma}{J_3}
 \cdot \aneufj{l_x'}{\sigma_x'}{j_x'}{l_y'}{\sigma_y'}{j_y'}{l_{xy}'}{S}{j_{xy}'}\\ 
&\cdot&\displaystyle \aneufj{l_{xy}}{\sigma}{J_3}{l_z}{s_4}{j_z}{L}{S}{J}
 \cdot (-)^{l_{xy}'+S-J_3'+L+S-J} 
 \cdot \asixj{S}{l_{xy}'}{J_3'}{l_z'}{J}{L}\\
&\cdot&\displaystyle \asixj{l_x'}{l_y'}{l_{xy}'}{l_z'}{L}{l_{yz}'}
 \cdot \asixj{l_x'}{\lambda}{l_{xy}}{l_z}{L}{l_{yz}'}
 \cdot (-)^{s_1+s_3-\sigma_x'+\sigma_x+s_3-\sigma}\\
&\cdot&\displaystyle \asixj{\sigma_x'}{s_2}{\sigma}{s_4}{S}{\sigma_y'}
\cdot \asixj{s_3}{s_1}{\sigma_x'}{s_2}{\sigma}{\sigma_x}
 \cdot \frac{x y z}{x^g_{\alpha \alpha'} y^g_{\alpha \alpha'} z^g_{\alpha \alpha'}} \\
&\cdot&\displaystyle {\cal{H}}^{l_{xy}}_{l_x l_y , l_x' \lambda}(x,y,u) 
\cdot {\cal{H}}^{l_{yz}'}_{\lambda l_z , l_y' l_z' }(y_0,z,v)
\end{eqnarray*}
where
$ \displaystyle t_{\alpha,\alpha'}  = 
(-)^{t_1+t_3-\tau_x'+\tau_x+t_3-T_3}
\cdot \asixj{\tau_x'}{t_2}{T_3}{t_4}{T}{\tau_y'}
\cdot \asixj{t_3}{t_1}{\tau_x'}{t_2}{\tau_y}{\tau_x}$
and with the constraints $\cos(\widehat{\hat{x},\hat{y}})=u$ and $\cos(\widehat{\hat{y}_0,\hat{z}})=v$.


\newpage

\begin{thebibliography}{10}

\bibitem{SSK92}
{N.~W.~Schellingerhout, J.~J.~Schut, L.~P.~Kok}.
\newblock {\em \prc{46}}, 4:1192, 1992.

\bibitem{GK93}
{W.~Gl\"{o}ckle, H.~Kamada}.
\newblock {\em \prl{71}}, 7:971, 1993.

\bibitem{GWKHG95}
{W.~Gl\"{o}ckle, H.~Witala, D.~H\"{u}ber, H.~Kamada, J.~Golak, Contrib. to the
  Int. Conf. Few-Body Problems in Physics, Williamsburg, may 1994, AIP
  Conference Proceedings 334, Ed. by F. Gross, pag. 45}.

\bibitem{PISA95}
{M.~Viviani, A.~Kievsky, S.~Rosati}.
\newblock {\em Few-Body Systems {18}}, page 25, 1995.

\bibitem{TJON76}
{J.~A.~Tjon}.
\newblock {\em Phys. Lett. B {63}}, 4:391, 1976.

\bibitem{F84}
{A.~C.~Fonseca}.
\newblock {\em \prc{30}}, 1:35, 1984.

\bibitem{F89}
{A.~C.~Fonseca}.
\newblock {\em \prc{40}}, 3:1390, 1989.

\bibitem{F94}
{A.~C.~Fonseca}.
\newblock {\em Few-Body Systems {Suppl.7}}, page 177, 1994.

\bibitem{UOT93}
{E.~Uzu, S.~Oryu, M.~Tanifuji}.
\newblock {\em Progress of Theor. Phys. {90}}, 4:937, 1993.

\bibitem{UOT95}
{E.~Uzu, S.~Oryu, M.~Tanifuji}.
\newblock {\em Few-Body Systems {Suppl.8}}, page~97, 1995.

\bibitem{APS91}
{A.~Arriaga, V.~R.~Pandharipande, R.~Schiavilla}.
\newblock {\em \prc{43}}, 3:983, 1991.

\bibitem{SCW95}
{R. Schiavilla, J. Carlson, R.B. Wiringa, Contrib. to the Int. Conf. Few-Body
  Problems in Physics, Williamsburg, may 1994, AIP Conference Proceedings 334,
  Ed. by F. Gross, pag. 79}.

\bibitem{YAKOVLEV95}
{S.~L.~Yakovlev, I.~N.~Filikhin}.
\newblock {\em Physics of Atomic Nuclei {58}}, 5:754, 1995.

\bibitem{HOF97}
{H.~M.~Hofmann, G.~M.~Hale}.
\newblock {\em Nucl. Phys. A {613}}, page 69, 1997.

\bibitem{CGM93}
{J.~Carbonell, C.~Gignoux, S.~P.~Merkuriev}.
\newblock {\em Few-Body Systems {15}}, page~15, 1993.

\bibitem{CCG97}
{F.~Ciesielski, J.~Carbonell, C.~Gignoux, Contrib. to the Int. Conf. Few-Body
  Problems in Physics, Groningen, july 1997}.

\bibitem{YAKU67}
{O.~A.~Yakubowsky}.
\newblock {\em Sov. J. Nucl. Phys. {5}}, page 937, 1967.

\bibitem{FAD60}
{L.~D.~Faddeev}.
\newblock {\em JETP}, 39:1459, 1960.

\bibitem{FAD61}
{L.~D.~Faddeev}.
\newblock {\em Sov. Phys. JETP}, 12:1014, 1961.

\bibitem{Fred97}
F.~Ciesielski, Th\`ese, Universit\'e J.~Fourier Grenoble (1997).

\bibitem{MERKU84}
{S.~P.~Merkuriev, S.~L.~Yakovlev, C.~Gignoux}.
\newblock {\em Nucl. Phys. A {431}}, page 125, 1984.

\bibitem{TAYLOR}
{J.~R.~Taylor}.
\newblock {\em Scattering Theory}.
\newblock Wiley \& Sons, Inc., 1972.

\bibitem{BENCZE95}
{Gy.~Bencze, C.~Chandler, A.~G.~Gibson, G.~W.~Pletsch}.
\newblock {\em Few-Body Systems {18}}, page 213, 1995.

\bibitem{PAYNE87}
{G.~L.~Payne}.
\newblock {\em Lecture Notes in Physics {93}}, page~64, 1987.

\bibitem{SKB89}
{N.~W.~Schellingerhout, L.~P.~Kok, G.~D.~Bosveld}.
\newblock {\em \pra{40}}, 10:5568, 1989.

\bibitem{SK90}
{N.~W.~Schellingerhout, L.~P.~Kok}.
\newblock {\em Nucl. Phys. A {508}}, page 290, 1990.

\bibitem{S95}
N.W. Schellingerhout, PhD, Groningen University (1995).

\bibitem{SAAD86}
{Y.~Saad, M.~H.~Schultz}.
\newblock {\em SIAM J. Sci. Stat. Comput. {7}}, page 856, 1986.

\bibitem{SCHELL89}
{N.~W.~Schellingerhout, L.~P.~Kok, G.~D.~Bosveld}.
\newblock {\em \pra{40}}, 10:5568, 1989.

\bibitem{MT69}
{R.~A.~Malfliet, J.~A.~Tjon}.
\newblock {\em Nucl. Phys. A {127}}, page 161, 1969.

\bibitem{SAAD}
{Y.~Saad}.
\newblock {\em Numerical Methods for Large Eigenvalue Problems}.
\newblock Manchester University Press Series in Algorithms and Architectures
  for Advanced Scientific Computing, New~York, 1992.

\bibitem{AZIZ3}
{R.~A.~Aziz, M.~J.~Slaman}.
\newblock {\em J. Chem. Phys. {94}}, 12:8047, 1991.

\bibitem{LA82}
{G.~L.~Payne, J.~L.~Friar, B.~F.~Gibson}.
\newblock {\em \prc{26}}, page 1385, 1982.

\bibitem{KG92}
{H.~Kamada, W.~Gl\"{o}ckle}.
\newblock {\em Nucl. Phys. A {548}}, page 205, 1992.

\bibitem{TILLEY92}
{D.~R.~Tilley, H.~R.~Weller, G.~.M.~Hale}.
\newblock {\em Nucl. Phys. A {541}}, page~1, 1992.

\bibitem{BLOM67}
{J.~Blomqvist}.
\newblock {\em Nucl. Phys. A {103}}, page 644, 1967.

\bibitem{WONG71}
{S.~K.~M.~Wong, G.~Saunier, B.~Rouben}.
\newblock {\em Nucl. Phys. A {169}}, page 294, 1971.

\bibitem{CVS88}
{R.~Ceuleneer, P.~Vandepeutte, C.~Semay}.
\newblock {\em \prc{38}}, 5:2335, 1988.

\bibitem{ADHIK81}
{S.~K.~Adhikari}.
\newblock {\em \prc{24}}, 1:16, 1981.

\bibitem{PHILL80}
{T.~W.~Phillips, B.~L.~Berman, J.~D.~Seagrave}.
\newblock {\em \prc{22}}, 2:384, 1980.

\bibitem{SEAG80}
{J.~D.~Seagrave, B.~L.~Berman, T.~W.~Phillips}.
\newblock {\em Phys. Lett. B {91}}, 2:200, 1980.

\bibitem{ISN7211}
A.~Laverne, C.~Gignoux, Rapport Interne ISN72.11.

\bibitem{RAUCH85}
{H.~Rauch, D.~Tuppinger, H.~W{\"{o}}lwitsch, T.~Wroblewski}.
\newblock {\em Phys. Lett. B {165}}, 1-2-3:39, 1985.

\bibitem{HALE90}
{G.~M.~Hale, D.~C.~Dodder, J.~D.~Seagrave, B.~L.~Berman, T.~W.~Phillips}.
\newblock {\em \prc{42}}, 1:438, 1990.

\end{thebibliography}


\newpage
\begin{table}[hbtp]
\narrowtext
\begin{tabular}{ccccccccccc}
\multicolumn{11}{c}{$S=0$ ($J^{\pi}=0^+$) \hspace{0.5cm} $T=0$} \\ \hline
K & $\tau_x$ & $T_3$ & $l_x$ & $\sigma_x$ & $j_x$ & $l_y$ & $j_y$ & $J_3^{\pi_3}$ & $l_z$ & $j_z$ \\
$\rightarrow$ &1&1/2&0&0&0&0&1/2&1/2$^+$&0&1/2 \\
$\rightarrow$ &0&1/2&0&1&1&0&1/2&1/2$^+$&0&1/2 \\ \hline
H & $\tau_x$ & $\tau_y$ & $l_x$ & $\sigma_x$ & $j_x^{\pi_x}$ & $l_y$ & $\sigma_y$ & $j_y^{\pi_y}$ & $j_{xy}$ & $l_z$ \\
&1&1&0&0&0$^+$&0&0&0$^+$&0&0 \\
$\sim$ &0&0&0&1&1$^+$&0&1&1$^+$&0&0 \\
\hline
\hline
\multicolumn{11}{c}{$S=1$ ($J^{\pi}=1^+$) \hspace{0.5cm} $T=0$} \\ \hline
K & $\tau_x$ & $T_3$ & $l_x$ & $\sigma_x$ & $j_x$ & $l_y$ & $j_y$ & $J_3^{\pi_3}$ & $l_z$ & $j_z$ \\
$\rightarrow$ &1&1/2&0&0&0&0&1/2&1/2$^+$&0&1/2 \\
$\rightarrow$ &0&1/2&0&1&1&0&1/2&1/2$^+$&0&1/2 \\
&0&1/2&0&1&1&0&1/2&3/2$^+$&0&1/2 \\ \hline
H & $\tau_x$ & $\tau_y$ & $l_x$ & $\sigma_x$ & $j_x^{\pi_x}$ & $l_y$ & $\sigma_y$ & $j_y^{\pi_y}$ & $j_{xy}$ & $l_z$ \\
&0&0&0&1&1$^+$&0&1&1$^+$&1&0 \\
\hline
\hline
\multicolumn{11}{c}{$S=0$ ($J^{\pi}=0^+$) \hspace{0.5cm} $T=1$} \\\hline
K & $\tau_x$ & $T_3$ & $l_x$ & $\sigma_x$ & $j_x$ & $l_y$ & $j_y$ & $J_3^{\pi_3}$ & $l_z$ & $j_z$ \\
$\rightarrow$ &1&1/2&0&0&0&0&1/2&1/2$^+$&0&1/2 \\
$\rightarrow$ &0&1/2&0&1&1&0&1/2&1/2$^+$&0&1/2 \\
&1&3/2&0&0&0&0&1/2&1/2$^+$&0&1/2 \\ \hline
H & $\tau_x$ & $\tau_y$ & $l_x$ & $\sigma_x$ & $j_x^{\pi_x}$ & $l_y$ & $\sigma_y$ & $j_y^{\pi_y}$ & $j_{xy}$ & $l_z$ \\
&1&1&0&0&0$^+$&0&0&0$^+$&0&0 \\
\hline
\hline
\multicolumn{11}{c}{$S=1$ ($J^{\pi}=1^+$) \hspace{0.5cm} $T=1$} \\\hline
K & $\tau_x$ & $T_3$ & $l_x$ & $\sigma_x$ & $j_x$ & $l_y$ & $j_y$ & $J_3^{\pi_3}$ & $l_z$ & $j_z$ \\
$\rightarrow$ &1&1/2&0&0&0&0&1/2&1/2$^+$&0&1/2 \\
$\rightarrow$ &0&1/2&0&1&1&0&1/2&1/2$^+$&0&1/2 \\
&1&3/2&0&0&0&0&1/2&1/2$^+$&0&1/2 \\
&0&1/2&0&1&1&0&1/2&3/2$^+$&0&1/2 \\ \hline
H & $\tau_x$ & $\tau_y$ & $l_x$ & $\sigma_x$ & $j_x^{\pi_x}$ & $l_y$ & $\sigma_y$ & $j_y^{\pi_y}$ & $j_{xy}$ & $l_z$ \\
&1&0&0&0&0$^+$&0&1&1$^+$&1&0 \\
&0&1&0&1&1$^+$&0&0&0$^+$&1&0 \\
\end{tabular}
\caption{Faddeev-Yakubovsky components for $(S,T)$ 4N states
in the S-wave approximation. The listed quantum numbers are those defined
by the coupling schemes (\ref{devcoupling}). The symbols `$\rightarrow$' and `$\sim$'
emphasize respectively the asymptotic N+NNN and NN+NN channels.}\label{tab_amplis}
\end{table}

\newpage
\begin{table}[hbt]
\narrowtext
\begin{tabular}{cddddd}  
        &  $^4$He \cite{SSK92} & $^4$He \cite{KG92} &  $^4$He & $^4$He$^*$ & $^3$H	\\  \hline
 B      &  30.31  	       &  30.29	          & 30.30   &  8.79      &	8.53  \\
r.m.s.  &	-		       &   - 	          &  1.44   &  4.95      &	1.72  \\
\end{tabular}
\caption{Binding energies (MeV) and r.m.s. radius (fm) for the 4N ground ($^4$He)
 and first excited ($^4$He$^*$) states. Our results for the $^4$He binding energy agree
 very well with the best existing calculations.
 The triton parameters are also mentioned for completeness.}\label{tab_bs}
\end{table}

\newpage
\begin{table}[hbt]
\narrowtext
\begin{tabular}{ccdddcdddcddd}
\multicolumn{13}{c}{grid G$_1$} \\ \hline
x && 20 & 1.30 & 10.0 && - & - & - &&-  &- &-	 \\
y && 15 & 1.25 & 12.0 && - & - & - &&-  &- &-	 \\
z && 15 & 1.25 & 15.0 && - & - & - &&-  &- &-	 \\
\hline
\multicolumn{13}{c}{grid G$_2$} \\ \hline
x && 15 & 1.30 & 10.0 && - & - & - &&-  &- &- \\
y && 10 & 1.25 & 12.0 && - & - & - &&-  &- &- \\
z && 10 & 1.25 & 15.0 && - & - & - &&-  &- &- \\
\hline
\multicolumn{13}{c}{grid G$^*_1$} \\ \hline
x && 08 & 1.30 & 08.0 && 05 & 1.10 & 20.0 && - & - & -  \\
y && 07 & 1.30 & 10.0 && 05 & 1.10 & 30.0 && - & - & -  \\
z && 07 & 1.20 & 10.0 && 13 & 1.10 & 80.0 && 10 & 1.00 & 150.0  \\
\end{tabular}
\caption{Grids used for 4N ground ($^4$He) and first excited ($^4$He$^*$) states.}\label{tab_grids1}
\end{table}

\newpage
\begin{table}[hbt]
\narrowtext
\begin{tabular}{clllddd} 
$N_K+N_H$ & $l_x$ & $l_y$   & $l_z$ & $B_4$      & $B_4^*$ &  $B^*-B_3$    \\ \hline  
  2 + 2   &   0   &   0     & 0     & 30.302 & 8.769 & 0.257    \\	
  8 + 2   &   0   &   0,1   & 0,1   & 30.319 & 8.763 & 0.250    \\	 
 16 + 3   &   0   &   0,1,2 & 0,1,2 & 30.324 & 8.770 & 0.257     \\   
\end{tabular}
\caption{Non zero angular momentum contributions to $^4$He and $^4$He$^*$ binding energies.}\label{tab_bs2}
\end{table}

\newpage
\begin{table}[hbt]
\narrowtext
\begin{tabular}{ccdddcddd}
\multicolumn{9}{c}{grid T$_4$~: $B_3=8.593$ MeV} \\ \hline 
x && 08 & 1.30 & 14.0 && 01 & - & 18.0  \\
y && 07 & 1.20 & 19.0 && 02 & 1.00 & 29.0  \\
\hline
\multicolumn{9}{c}{grid T$_8$~: $B_3=8.527$ MeV} \\ \hline 
x && 12 & 1.30 & 14.0 && 01 & -    & 18.0  \\
y && 10 & 1.20 & 19.0 && 02 & 1.00 & 29.0  \\
\hline
\multicolumn{9}{c}{grid T$_{10}$~: $B_3=8.535$ MeV} \\ \hline 
x && 18 & 1.20 & 14.0 && 02 & 1.00 & 18.0  \\
y && 15 & 1.10 & 19.0 && 04 & 1.00 & 29.0  \\
\end{tabular}
\caption{The grids T$_4$, T$_8$, T$_{10}$ used for the tritons 4, 8, 10.}\label{grids2}
\end{table}

\newpage
\begin{table}[hbt]
\narrowtext
\begin{tabular}{ddd}
$E_c$~(MeV) & \multicolumn{2}{c}{$\delta \,\, (^{\circ})$} \\ \hline
{\boldmath $S=0, T=1$}&T$_4$ &T$_8$ \\ \hline
0.05 &169.97&169.95 \\
0.1  &165.85&- \\
0.5  &148.99&- \\
1.0  &137.13&137.07 \\
2.0  &121.79&- \\
3.0  &111.27&- \\
4.0  &102.93&102.82 \\
5.0  & 96.18 &96.13 \\
6.0  & 90.18 &90.24 \\
\hline
{\boldmath $S=0, T=0$}&T$_4$ &T$_8$ \\
\hline
0.01&165.17&164.05 \\
0.02&159.17&157.66 \\
0.05&147.73&-	  \\
0.07&142.29&-	  \\
0.1 &135.72&132.98 \\
0.2 &120.52&-      \\
0.3 &109.52     &106.95 \\
0.5 &95.81 &-	  \\
0.7 &85.57 &-	  \\
1.0 &74.08 &71.05  \\
2.0 &49.92 &47.16  \\
3.0 &34.71 &32.07  \\
4.0 &23.60 &21.04  \\
\hline
{\boldmath $S=1, T=1$}&& T$_8$ \\ \hline
0.03&&172.97 \\
0.06&&170.07 \\
0.12&&166.0  \\
0.3 &&158.02 \\
0.9 &&142.84 \\
1.8 &&129.11 \\
2.7 &&119.43 \\
3.6 &&111.84 \\
5.4 &&100.10 \\
6.3 &&94.84  \\
\hline
{\boldmath $S=1, T=0$}&& T$_8$ \\ \hline
0.01&&176.46 \\
0.02&&174.99 \\
0.05&&172.09 \\
0.1 &&168.83 \\
0.5 &&155.23 \\
1.0 &&145.35 \\
2.0 &&132.06 \\
3.0 &&122.53 \\
4.0 &&114.96 \\
5.0 &&108.68 \\
6.0 &&103.24 \\
\end{tabular}
\caption{N+NNN phase-shifts (degrees), as a function of the center of mass kinetic energy
(MeV), in different $(S,T)$ channels. In the $S=0$ case, the results with
different grids are shown.}\label{DEPHAN_NNN}
\end{table}

\newpage
\begin{table}[hbt]
\narrowtext
\begin{tabular}{ccdddcd}
\multicolumn{6}{c}{$a_{ST}$~(fm) }\\ \hline
S & T &triton~8&triton~10&\multicolumn{3}{c}{final value}  \\ \hline
0 & 1 & 4.13 & 4.13&  4.13 &$\pm$& 0.01  \\
1 & 1 & 3.73 & 3.73&  3.73 &$\pm$& 0.01  \\
0 & 0 &14.78 &14.76& 14.76 &$\pm$& 0.02  \\
1 & 0 & 3.25 & 3.25&  3.25 &$\pm$& 0.01  \\
\end{tabular} 
\caption{N+NNN scattering lengths values in different (S,T) channels,
in the S-wave approximation.}\label{tab_sl}
\end{table}

\newpage
\begin{table}[hbt]
\narrowtext
\begin{tabular}{ccddddd}
S&T& $a$ (fm)     &  $r_0$ (fm) & $v_0$ (fm$^3$) & $q_0$ (fm$^{-1})$ &  $a$ \cite{YAKOVLEV95}  \\ \hline
0&1& 14.75 &  6.75  & 0.462        & -	             &          \\
1&1&  3.25 &  1.82  & 0.231        & -	             &          \\
0&0&  4.13 &  2.01  & 0.308        & 0.505	     &  4.0             \\
1&0&  3.73 &  1.87  & $\simeq$ 0   & -               &  3.6             \\
\end{tabular}
\caption{Low energy N+NNN parameters, in the S-wave approximation.}\label{tab_lep}
\end{table}

\newpage
\begin{table}[hbt]
\narrowtext
\begin{tabular}{cc|cc|cc|cc|cc}
 S&T&\multicolumn{2}{c|}{$l_y,l_z=0$}&\multicolumn{2}{c|}{$l_y,l_z=0,1$}
    &\multicolumn{2}{c|}{$l_y,l_z=0,1,2$}&\multicolumn{2}{c}{$l_y,l_z=0,1,2,3$}\\\hline
  & &      $a$   &$N_c$&       $a$  &$N_c$&       $a$  &$N_c$ &  $a$     &$N_c$  \\
 0&0&  14.78   &4&    14.86 &10&   14.72 &16& 14.72  &22\\
 1&0&	3.25   &4&     3.08 &17&    3.08 &31&  3.08  &45\\
 0&1&	4.13   &4&     4.10 &12&    4.10 &20&  4.10  &28\\
 1&1&	3.73   &6&     3.63 &23&    3.63 &41&  3.63  &59\\
\end{tabular}
\caption{Convergence of low energy N+NNN scattering lengths,
with respect to increasing internal angular momenta.}\label{ldifl}
\end{table}

\newpage
\begin{table}[hbt]
\widetext
\begin{tabular}{ddcccc}
\multicolumn{2}{c}{$E_c\;$ c.m. (MeV)}&\multicolumn{4}{c}{S-matrix}\\ \hline
N+NNN&NN+NN&$S_{11}$&$S_{12}$&$S_{21}$&$S_{22}$\\ \hline
4.1253&0.05&	0.772+0.634i	&-0.00473+0.0472i &		-0.00471+0.0470i	&0.882-0.468i	\\
4.1753&0.10&	0.782+0.621i	&0.000488+0.0564i &		0.000486+0.0563i	&0.771-0.634i	\\
4.2753&0.20&	0.802+0.594i	&0.0110+0.0661i &		0.0110+0.0661i	&0.566-0.821i   \\
4.3253&0.25&	0.811+0.581i	&0.0160+0.0690i &		0.0161+0.0691i	&0.472-0.879i   \\
4.3753&0.30&	0.820+0.567i	&0.0209+0.0711i &		0.0210+0.0713i	&0.382-0.921i   \\
4.5753&0.50&	0.854+0.514i	&0.0391+0.0748i &		0.0393+0.0752i	&0.0659-0.994i  \\
5.0753&1.00&	0.919+0.381i	&0.0759+0.0679i &		0.0758+0.0678i	&-0.480-0.871i  \\
5.5753&1.50&	0.961+0.253i	&0.103+0.0508i &		0.102+0.0506i	&-0.784-0.610i \\
6.0753&2.00&	0.983+0.133i	&0.120+0.0292i &		0.123+0.0301i	&-0.934-0.335i \\
\end{tabular}
\caption{N+NNN$\leftrightarrow$NN+NN S-matrix elements. The N+NNN channel is labeled by 1,
the NN+NN one by 2.}\label{Smat}
\end{table}

\newpage
\begin{table}[hbt]
\narrowtext
\begin{tabular}{dcddcdc}
\multicolumn{3}{c}{$a_0$} & \multicolumn{3}{c}{$a_1$} \\ \hline
3.91 &$\pm$&0.12  & 3.6  &$\pm$&0.1   &\cite{PHILL80}    \\
4.98 &$\pm$&0.29  & 3.13 &$\pm$&0.11  &\cite{RAUCH85}    \\
2.10 &$\pm$&0.31  & 4.05 &$\pm$&0.09  &\cite{RAUCH85}    \\
4.453&$\pm$&0.10  & 3.325&$\pm$&0.016 &\cite{HALE90}     \\
 \end{tabular}
\caption{Latest experimental results concerning
n+t singlet ($a_0$) and triplet ($a_1$) scattering lengths (in fm).}\label{tab3}
\end{table}

\widetext

\newpage

\newpage

\newpage

\newpage

\newpage

\newpage

\newpage

\newpage

\newpage

\newpage

\end{document}